%% file: iclr2024_conference.tex
\documentclass{article} % For LaTeX2e
\usepackage{iclr2024_conference,times}

\input{math_commands.tex}

\usepackage{hyperref}
\usepackage{url}

\usepackage{graphicx}
\usepackage{subcaption}
\usepackage{booktabs}
\usepackage{multirow}
\usepackage{adjustbox}

\title{Uncertainty Aware Tropical Cyclone Wind Speed Estimation from Satellite Data}

\author{Nils Lehmann \\
Technical University of Munich\\
{\tt\small n.lehmann@tum.de}
\And
Nina Maria Gottschling\\
EO Data Science, DLR\\
{\tt\small nina-maria.gottschling@dlr.de}
\And
Stefan Depeweg\\
Siemens AG\\
{\tt\small stefan.depeweg@siemens.com}
\And
Eric Nalisnick\\
University of Amsterdam\\
{\tt\small e.t.nalisnick@uva.nl}
}

\iclrfinalcopy % Uncomment for camera-ready version, but NOT for submission.
\begin{document}

\maketitle

\begin{abstract}
Deep neural networks (DNNs) have been successfully applied to earth observation (EO) data and opened new research avenues. Despite the theoretical and practical advances of these techniques, DNNs are still considered black box tools and by default are designed to give point predictions. However, the majority of EO applications demand reliable uncertainty estimates that can support practitioners in critical decision making tasks. This work provides a theoretical and quantitative comparison of existing uncertainty quantification methods for DNNs applied to the task of wind speed estimation in satellite imagery of tropical cyclones. We provide a detailed evaluation of predictive uncertainty estimates from state-of-the-art uncertainty quantification (UQ) methods for DNNs. We find that predictive uncertainties can be utilized to further improve accuracy and analyze the predictive uncertainties of different methods across storm categories. 

\end{abstract}
\vspace{-10pt}
\section{Introduction}
The tremendous success of Deep Learning approaches to natural images is increasingly being explored on EO data that is becoming available in ever greater quantities \citep{tuia2023artificial}. Due to their often vast global coverage, EO data is an indispensable source of information for assessing the state of our planet as well as extreme events that are increasing in frequency and intensity \citep{kikstra2022ipcc}. One category of such extreme events are tropical cyclones. Tropical cyclones - in the US alone - have lead to 6,789 deaths and caused financial damages amounting to a staggering \$1,333.6 billion between 1980-2022, with an average instance cost of \$22.2 billion and covering 53.9\% of all costs caused by US extreme weather disasters \citep{noaaDamageDatabase}. Although, satellite data and other in-situ measurements are often available, reliable wind speed estimation remains a challenging task. For example in October 2023, hurricane Otis underwent a rapid intensification of almost 80 kts in 12 hours before causing devastating damage in the city Acapulco.\footnote{\href{https://www.nesdis.noaa.gov/news/hurricane-otis-causes-catastrophic-damage-acapulco-mexico}{"Hurricane Otis Causes Catastrophic Damage in Acapulco, Mexico", NOAA} accessed 31.01.2024.} The failure of satellite based wind speed estimation methods \citep{kramer2023daily} and the need for improving these has been highlighted after this tropical cyclone\footnote{\href{https://www.science.org/content/article/hurricane-otis-smashed-mexico-and-broke-records-why-did-no-one-see-it-coming}{"Hurricane Otis smashed into Mexico and broke records. Why did no one see it coming?"} accessed 31.01.2024.}. Moreover, rapidly intensifying storms near coastlines have shown a trend to become more frequent \citep{li2023recent} and, hence, this demonstrates the need for improved monitoring of wind speeds and better prediction methods to yield improved warning systems. Because data to train such prediction methods can be limited and unevenly distributed, making a perfect prediction is not always possible. However, based on the general viability of DNNs for predicting and estimating wind speeds from satellite data (see e.g. \cite{pradhan2017tropical}), one possible approach is to equip DNNs with modern uncertainty-quantification (UQ) methods to enhance the quality of predictions and mitigate data imbalances, as well as label and input noise. This uncertainty is important for EO applications, as in practice, a prediction model is only an element of a complex decision making process. For instance, the confidence in the prediction of a tropical cyclone category is a key factor for deciding on public safety measures. This paper has the following contribution:
 Using the dataset proposed in \cite{tropicalcyclone}, we show that equipping DNNs with predictive uncertainty can be utilized to further improve accuracy via selective prediction based on predictive uncertainty. To the best of our knowledge no previous related work (see Section \ref{sec:related_work}) considered an evaluation of uncertainty aware regression models in this domain. We compare state-of-the-art UQ methods, (see Section \ref{sec:methods}), and demonstrate differences across storm categories according to the Saffir-Simpson scale and different dataset splits. We show that UQ can improve real-time wind speed estimation and thus outline the way to apply UQ to DNN forecasting models by a detailed assessment of existing UQ methods.

\subsection{Related Work}\label{sec:related_work}

Several works have tackled the task of applying Deep Learning methods to tropical cyclone intensity estimation as a classification \citep{wimmers2019using} or regression \citep{chen2019estimating, ma2024multi, zhang2021tropical} task. Based on a dataset of 25k images of infrared satellite imagery matched with storm data from the HURDAT2 database \citep{landsea2013atlantic}, \cite{pradhan2017tropical} train a CNN architecture for storm-category classification, as well es wind-speed estimation, and demonstrate improvements over previously applied statistical techniques like Advanced Dvorak Technique (ADT) \citep{pineros2011estimating}, and Deviation-Angle Variance Technique (DAVT) \citep{ritchie2014satellite}. \cite{maskey2020deepti} improve the dataset quality and size by using GEOS Geostationary Operational Environmental Satellite (GEOS) and demonstrate a live production system. 
Our work is mostly comparable to \cite{maskey2020deepti} as we use their published dataset that was part of the Driven Data Challenge \citep{tropicalcyclone}.

\section{Tropical Cyclone Dataset}\label{sec:tropical_cyclone_dataset}
 \vspace{-5pt}
\begin{table}[h!]
\centering
\resizebox{\columnwidth}{!}{%
\begin{tabular}{@{}l|llllll@{}}
\toprule
Dataset name        & Satellite & Spatial Res & Temporal  Res & Train Samples & Val Samples & Test Samples \\ \midrule
Tropical Cyclone & GOES      & 2km         & 15 min        & 53k           & 11k         & 43k          \\
\bottomrule
\end{tabular}%
}
\caption{Dataset Overview}
\label{tab:dataset_overview}
\end{table}

\begin{figure}[h!]
     \centering
     \begin{subfigure}[b]{0.45\textwidth}
         \centering
         \includegraphics[width=\textwidth]{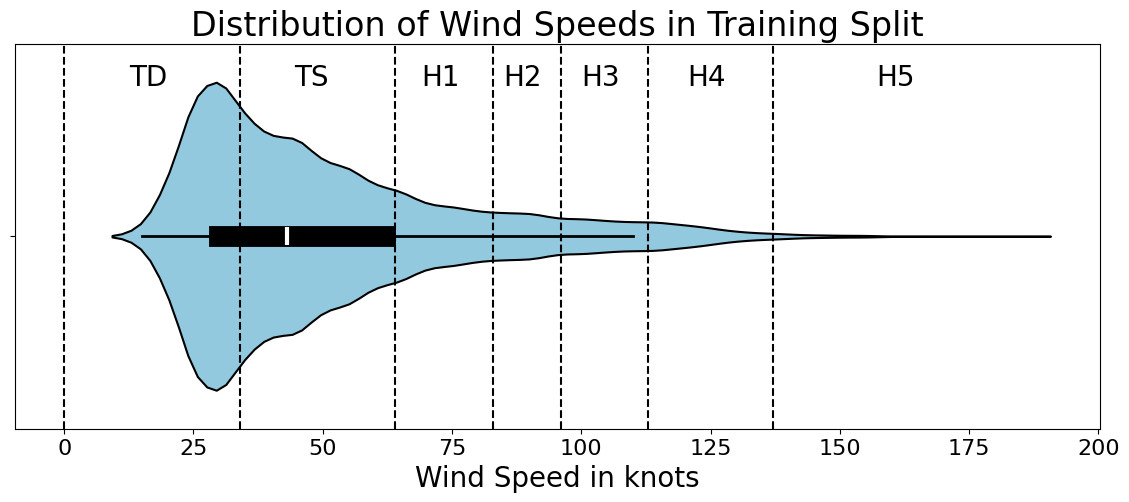}
         \caption{Label distribution and storm categories.}
         \label{fig:label_dist}
     \end{subfigure}
     \hfill
     \begin{subfigure}[b]{0.5\textwidth}
         \centering
         \includegraphics[width=\textwidth]{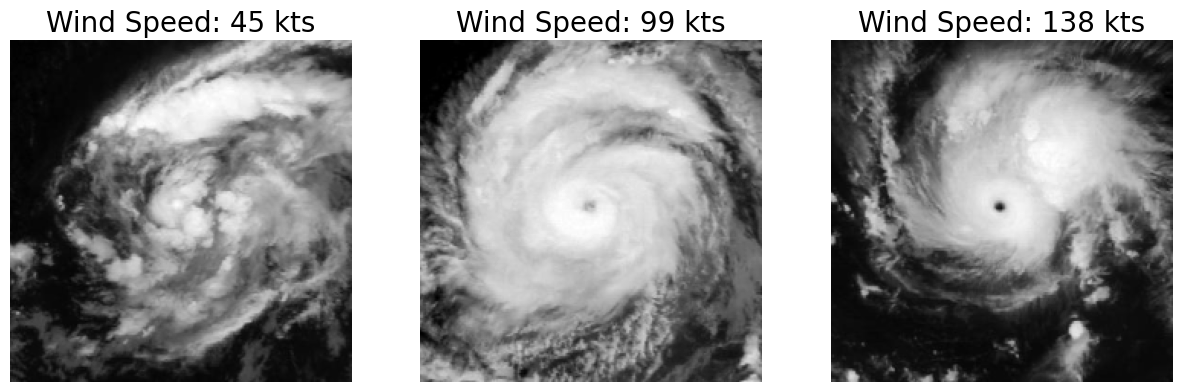}
         \vspace*{1mm}
         \caption{Dataset samples}
         \label{fig:image_examples}
     \end{subfigure}
\caption{ Visualization of Tropical Cyclone Dataset.}
\label{fig:dataset}
\end{figure}

\vspace{-5pt}
The imagery represents single channel long-wave infrared measurements captured every 15 minutes, at 10.3 microns, that can capture the spatial structure of the storm in terms of measurements of the brightness temperature, as seen in Figure \ref{fig:image_examples}. For more details about dataset collection, we refer the reader to the methodology section of \citep{maskey2020deepti}. We resize the images to 224x224 pixels and employ common image augmentations during training. We follow the datasplits by storm of the challenge and use dataloading available through the TorchGeo library \citep{stewart2022torchgeo}, which yields 53k training, 11k validation and 43k test samples. As Figure \ref{fig:label_dist} shows, the distribution of targets is highly skewed with the majority of samples falling beneath hurricane categories defined by the Saffir Simpson Scale, \cite{simpson1974hurricane}. We conduct experiments with the full target range but also subsets that only contain hurricane categories.

\vspace{-5pt}

\vspace{-5pt}

\section{Methods}\label{sec:methods}

\vspace{-5pt}

Given a set of input-target pairs $\mathcal{D}_{\mathrm{train}} = \{(x_i, y_i)\}_{i=1}^{N}$, $(x_i, y_i)$, the task of the neural network is to predict a target $y^\star \in Y$ given an input $x^\star \in X$. The input is a triplet of monochrome satellite images at time steps $[t-2, t-1, t]$ and the target is the maximum sustained wind speed in knots (kts) at time step $t$. \footnote{We choose this input image composition, as it was utilized in the winning solution of the challenge \citep{tropicalcyclone}, which improved reported accuracy significantly compared to \citep{maskey2020deepti}.} This is sometimes referred to as "nowcasting". For this task, we compare five classes of UQ methods: deterministic, ensemble, Bayesian, quantile and diffusion based methods. Firstly, deterministic UQ methods use a DNN, $f_{\theta}: X \rightarrow \mathcal{P}(Y)$, that map inputs $x$ to the parameters of a probability distribution $f_{\theta}(x^\star)=p_{\theta}(x^\star) \in \mathcal{P}(Y)$. These include Deep Evidential Networks (\textbf{DER}) \cite{amini2020deep}, where we use the correction proposed by \cite{meinert2023unreasonable}, and Mean Variance Networks (\textbf{MVE}) \citep{nix1994estimating} which output the mean and standard deviation of a Gaussian distribution $f_{\theta}^{\text{MVE}}(x^\star)=(\mu_{\theta}(x^\star),\sigma_{\theta}(x^\star))$. Secondly, the broadly considered state-of-the-art method Deep Ensembles (\textbf{DeepEnsembles}) proposed by \cite{lakshminarayanan2017simple} utilizes an ensemble over MVE networks. Thirdly, Bayesian methods aim at modelling a distribution over the network parameters and are commonly used to approximate the first and second moment of a marginalized distribution. These include Bayesian Neural Networks with Variational Inference (\textbf{BNN VI ELBO}) \cite{blundell2015weight}, MC-Dropout (\textbf{MCDropout}) \cite{gal2016dropout}, the Laplace Approximation (\textbf{Laplace}) \cite{ritter2018scalable}\cite{daxberger2021laplace} and \textbf{SWAG} \cite{maddox2019simple} with partially stochastic variants presented in \cite{sharma2023bayesian}. Gaussian Process based methods model a distribution over functions that also approximate the fist and second moment of the marginalized distribution. These include Deep Kernel Learning (\textbf{DKL}) \cite{wilson2016deep} and an extension thereof Deterministic Uncertainty Estimation (\textbf{DUE}) \citep{van2021feature}. Fourthly, quantile based models $f_{\theta}: X \rightarrow Y^n$ that map to $n$ quantiles, $f_{\theta}(x^\star) = (q_1(x^\star),...,q_n(x^\star)) \in Y^n$, such as Quantile Regression (\textbf{Quantile Regression}) and the conformalized version thereof (\textbf{ConformalQR}) suggested by \cite{romano2019conformalized}. Lastly, we also consider a diffusion model (\textbf{CARD}) as introduced by \cite{han2022card}. A detailed description of the methods is provided in the supplementary material. 
Depending on underlying assumptions UQ, methods are regarded to express two different types of uncertainties \citep{hullermeier2021aleatoric}. \textbf{Aleatoric uncertainty} refers to inherent randomness in the data and \textbf{epistemic uncertainty} to a lack of knowledge in the modelling process. From a statistical perspective \cite{gruber2023sources} allude that such a distinction is often not possible. Thus, we focus solely on predictive uncertainty.

\textbf{Evaluation methodology:}\label{sec:experiments} In addition to standard metrics for regression, such as root-mean-squared error (RMSE), we utilize proper scoring rules such as the negative log-likelihood (NLL) and continuous ranked probability score (CRPS) \citep{gneiting2007strictly} and the mean absolute calibration error (MACE). To evaluate the merit of UQ methods for decision making, we use selective prediction as a downstream task. Here, samples with a predictive uncertainty above a given threshold are omitted from prediction and can be referred to an expert or other estimation methods. Based on the Saffir-Simpson Scale \citep{simpson1974hurricane} bin intervals, we chose the threshold such that it would on average shift the category of the regression prediction. Hence, we take the threshold to be the mean over categories of the wind speed interval from categories $1$ to $4$, which is approximately $9$ kts. We experiment with different threshold choices which are reported in the supplementary material and in Fig. \ref{fig:due}. All methods have an ImageNet pretrained ResNet-18 \citep{he2016deep} backbone available frome the timm library \citep{rw2019timm}. Metrics are computed with the UQ-toolbox by \cite{chung2021uncertainty}.\footnote{Code available under \text{https://github.com/nilsleh/tropical\_cyclone\_uq}}

\begin{table}[h!]
\centering
\resizebox{10cm}{!}{%
\begin{tabular}{llrrrrrr}
\toprule
UQ group & Method & RMSE $\downarrow$ & RMSE $\Delta$ $\uparrow$ & Coverage $\uparrow$ & CRPS $\downarrow$ & NLL $\downarrow$ & MACE $\downarrow$ \\
\midrule
None & Deterministic & 10.50 & 0.00 & 1.00 & NaN & NaN & NaN \\
\midrule
\multirow{2}{*}{Deterministic} & \textbf{MVE} & 9.95 & 2.10 & \textbf{0.62} & \textbf{5.31} & \textbf{3.64} & 0.04 \\
 & DER & 10.14 & NaN & 0.00 & 10.07 & 4.60 & 0.35 \\
\midrule
\multirow{2}{*}{Quantile} & QR & 10.95 & 3.28 & 0.44 & 5.82 & 3.73 & \textbf{0.01} \\
 & CQR & 10.95 & \textbf{6.18} & 0.08 & 5.98 & 3.79 & 0.10 \\
\midrule
\multirow{1}{*}{Ensemble} & Deep Ensemble & 16.19 & 0.00 & 0.00 & 8.83 & 4.15 & 0.05 \\
\midrule
\multirow{6}{*}{Bayesian} & MC Dropout & 10.23 & 6.12 & 0.00 & 5.78 & 3.81 & 0.16 \\
 & \textbf{SWAG} & 9.78 & 5.42 & 0.11 & 5.40 & 3.71 & 0.13 \\
 & Laplace & 10.53 & 0.00 & 0.00 & 7.96 & 4.31 & 0.28 \\
 & BNN VI ELBO & \textbf{9.27} & 0.00 & 1.00 & 6.28 & 52.60 & 0.41 \\
 & DKL & 12.59 & 0.00 & 0.00 & 6.84 & 3.95 & 0.06 \\
 & DUE & 9.95 & 0.00 & 0.00 & 5.43 & 3.73 & 0.08 \\
\midrule
\multirow{1}{*}{Diffusion} & CARD & 10.86 & 1.50 & 0.60 & 5.84 & 3.92 & 0.05 \\
\bottomrule
\end{tabular}%
}
\caption{Evaluation Results on test set. RMSE $\Delta$ shows the improvement after selective prediction, where $0.00$ indicates that all samples were withdrawn, while Coverage denotes the fraction of remaining samples that were not omitted. Threshold 9 kts.}
\label{tab:results}
\end{table}

\vspace{-5pt}

%\begin{table}[h!]
%\centering
%\resizebox{10cm}{!}{%
%\begin{tabular}{llrrrrrr}
%\toprule
%UQ group & Method & RMSE $\downarrow$ & RMSE $\Delta$ $\uparrow$ & Coverage $\uparrow$ & CRPS $\downarrow$ & NLL $\downarrow$ & MACE $\downarrow$ \\
%\midrule
%None & Deterministic & 10.50 & 0.00 & 1.00 & NaN & NaN & NaN \\
%\midrule
%\multirow{2}{*}{Deterministic} & MVE & 9.95 & 2.10 & 0.62 & \textbf{5.31} & \textbf{3.64} & 0.04 \\
% & DER & 10.14 & NaN & 0.00 & 10.07 & 4.60 & 0.35 \\
%\midrule
%\multirow{2}{*}{Quantile} & QR & 10.95 & 3.28 & 0.44 & 5.82 & 3.73 & \textbf{0.01} \\
% & CQR & 10.95 & \textbf{6.18} & 0.08 & 5.98 & 3.79 & 0.10 \\
%\midrule
%\multirow{1}{*}{Ensemble} & Deep Ensemble & 16.19 & NaN & 0.00 & 8.83 & 4.15 & 0.05 \\
%\midrule
%\multirow{6}{*}{Bayesian} & MC Dropout & 10.23 & 6.12 & 0.00 & 5.78 & 3.81 & 0.16 \\
% & SWAG & \textbf{9.78} & 5.42 & 0.11 & 5.40 & 3.71 & 0.13 \\
% & Laplace & 10.53 & NaN & 0.00 & 7.96 & 4.31 & 0.28 \\
% & BNN VI ELBO & 11.82 & 0.55 & \textbf{0.95} & 6.70 & 5.57 & 0.23 \\
% & DKL & 12.59 & NaN & 0.00 & 6.84 & 3.95 & 0.06 \\
% & DUE & 9.95 & NaN & 0.00 & 5.43 & 3.73 & 0.08 \\
%\midrule
%\multirow{1}{*}{Diffusion} & CARD & 10.86 & 1.50 & 0.60 & 5.84 & 3.92 & 0.05 \\
%\bottomrule
%\end{tabular}%
%}
%\caption{Evaluation Results on test set. RMSE $\Delta$ shows the improvement after selective prediction, while Coverage denotes the fraction of remaining samples that were not omitted. Threshold 9 knots.}
%\label{tab:results}
%\end{table}

\vspace{-5pt}
\section{Results}\label{sec:results}
We show fine grained results for storm categories Tropical Depression (TD), and Hurricane categories 1, 3, and 5 for better visualization. Additional results for different dataset splits and thresholds including all categories are included in the supplementary material.

\begin{figure}[h!]
\vspace{-8pt}
     \centering
        \includegraphics[width=14cm,height=7cm,keepaspectratio]{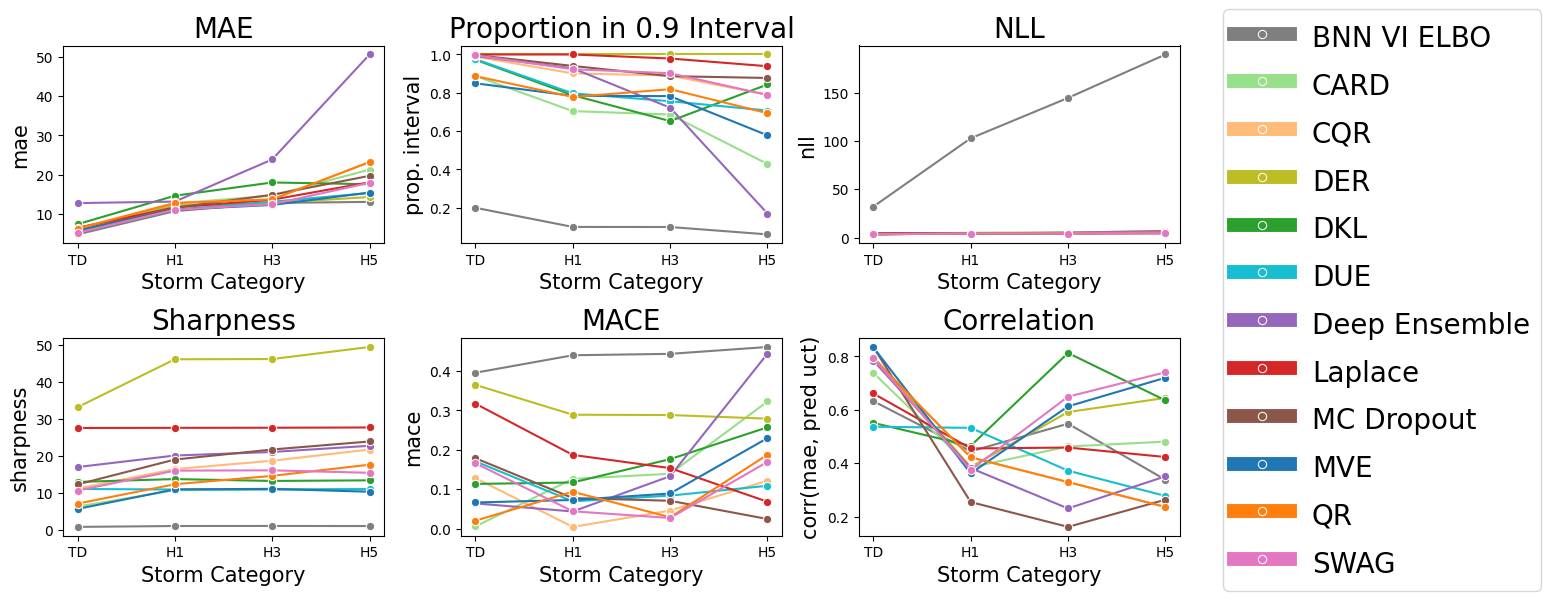}
        \caption{Uncertainty Metrics over different storm categories. We find that VI BNNs under cover (e.g. see proportion in interval), DER tends to over cover (e.g. see sharpness), with many other methods performing in between.}
        \label{fig:uq_metrics_result}
        \vspace{-5pt}
\end{figure}

\textbf{How effective is selective prediction?}
As Table \ref{tab:results} shows, selective prediction - enabled through uncertainy aware models - can yield significant accuracy improvements for selected methods. The best performing methods obtain an RMSE between $9.27-10.95$ kts, yet the accuracy improvement obtained by selective prediction varies significantly. However, the coverage - the remaining samples after selective prediction - also varies considerably.
%We attempted to inform our selective prediction threshold by practical considerations, however, there might be better choices that would improve results.
For higher hurricane categories, accuracy and uncertainty metrics worsen substantially as shown in Figure \ref{fig:uq_metrics_result} and different ranges of improvement are obtained by selective prediction, as shown in the supplementary material. %On higher categories SWAG, Laplace and CQR obtain relatively low calibration errors, MACE, while at the same time having relatively low MAE and a large proportion in the 0.9 interval. 
When averaging over all categories Table \ref{tab:results} shows that SWAG and CQR obtain relatively low RMSE after seletive prediction, 4.36 and 4.77 kts, while maintaining a coverage of 11\% and 8\%. 

\begin{figure}[h!]
     \centering
     \begin{subfigure}[b]{0.49\textwidth}
        \centering
        \includegraphics[width=\textwidth]{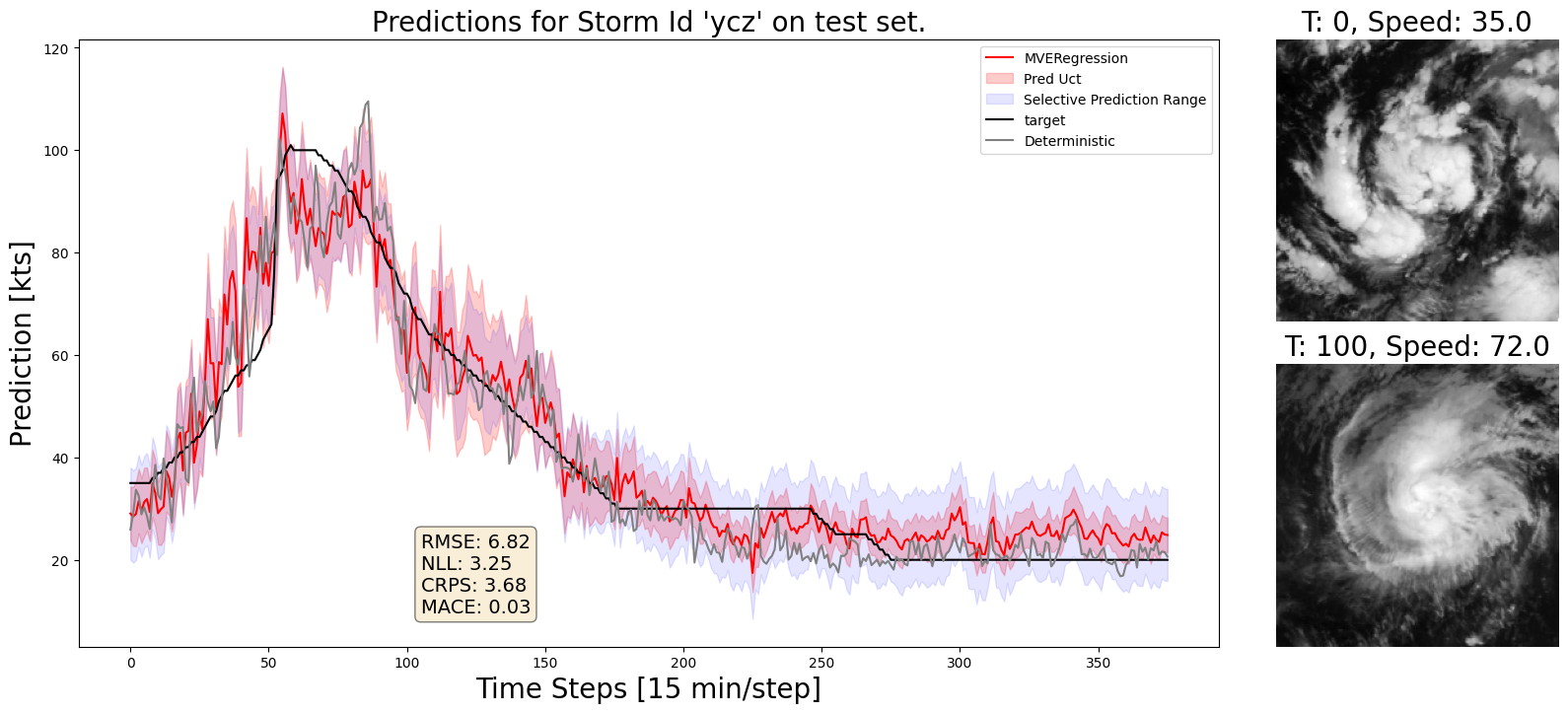}
        \caption{MVE prediction Example with a visualized threshold of 9 kts.}
        \label{fig:mve}
     \end{subfigure}
     \hfill
     \begin{subfigure}[b]{0.49\textwidth}
        \centering
        \includegraphics[width=\textwidth]{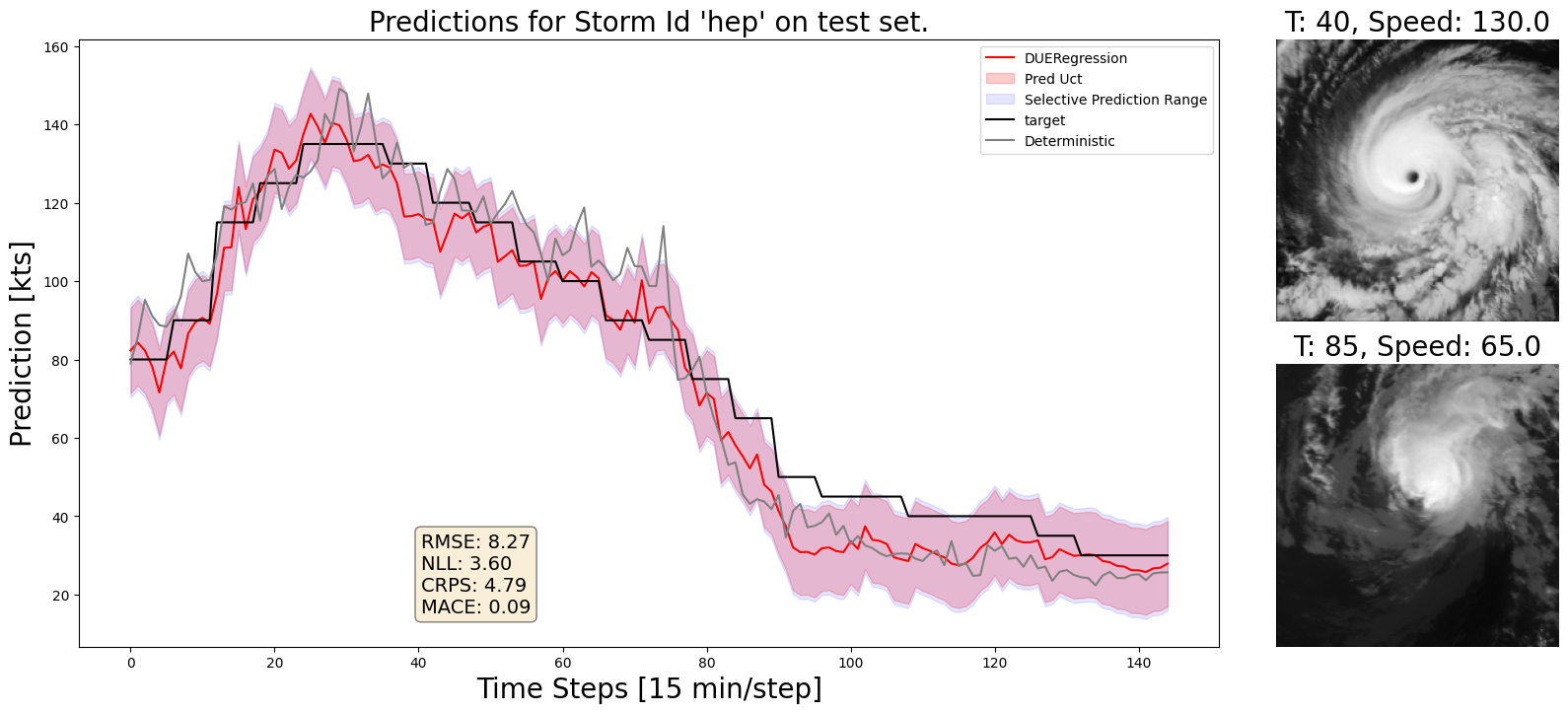}
        \caption{DUE Prediction Example with a visualized threshold of 12 kts.}
        \label{fig:due}
     \end{subfigure}
\caption{Predictive Uncertainty Examples. Note that models under our setup do not have a concept of time, we merely combine individual nowcasting predictions into a time-series. Red shaded areas exceeding blue areas indicate samples that \textit{would} be omitted during selective prediction. Figure inspired by \cite{zhang2019tropical}.}
\label{fig:result_2}
\end{figure}

\textbf{Error and Predictive Uncertainty across Categories:} We evaluate the predictive uncertainty across storm categories with three criteria: the correlation between predictive uncertainty and MAE, sharpness, and MACE. Fig. \ref{fig:uq_metrics_result}, on the bottom right, shows the correlation is best for the TD case and is fairly consistent for most models. On the higher categories H3 and H5 we observe a larger spread between models, with SWAG, MVE, DKL and DER demonstrating higher correlation values. Accurate predictive uncertainties need to be both well calibrated - obtain a low MACE - and be sharp \cite{kuleshov2018accurate}. MVE, SWAG, QR and CQR most closely fulfill this criteria. In contrast, Laplace and MC-Dropout obtain a low MACE on higher categories but are also less sharp and show lower correlation. Fig. \ref{fig:mve} gives a "qualitative" example of MVE predictions for a selected storm track which generally follows the trend of the underlying target. Samples with a predictive uncertainty that exceeds the selective prediction threshold could be referred to an expert or postprocessing step.

\vspace{-5pt}
\vspace{-10pt}

\subsection{Detailed discussion per UQ method group}\label{sec:discussion}
\vspace{-5pt}
\textbf{Deterministic UQ methods}: Table \ref{tab:results} shows MVE obtains an RMSE of 7.85 kts after selective prediction while maintaining a coverage of 62\% and the lowest scoring rules, NLL and CRPS, which may be correlated to the fact that the loss objective is the NLL. At the same time MVE remains well calibrated compared to all other methods. Table 1 in the Appendix, Section 1, shows that MVE also obtains a comparably low RMSE and NLL per category. DER obtains a higher RMSE and no improvement with selective prediction, Table \ref{tab:results}. This may be due to the fact that the predicted standard deviations of DER are relatively high compared to the selective prediction threshold, which is reflected in the sharpness accross storm categories in Figure \ref{fig:uq_metrics_result}.
\textbf{Quantile based UQ methods}: CQR obtains higher improvements with selective prediction than QR, see Table \ref{tab:results}, which is due to conformalization of quantiles and the resulting shift in predictive uncertainty. Yet this comes at the cost of a significantly lower coverage of CQR after selective prediction with only 8\% compared to 44\% for QR. 
\textbf{Ensemble methods}: Table \ref{tab:results} shows that overall Deep Ensembles obtain a higher RMSE than all other methods and also a significantly higher RMSE on category 5 cyclones, see Figure \ref{fig:uq_metrics_result}. Although Deep Ensembles are considered state-of-the-art, \cite{seligmann2024beyond} also find that they do not perform best at every UQ task. As for DER the predictive uncertainty of Deep Ensembles is larger than the selective prediction threshold, resulting in no improvement in RMSE. However, choosing a different threshold may result in accuracy improvements. We hypothesize that the variance of ensemble members might not be large enough and instead have converged to similar solutions, which implies that the ensemble members have similar biases.
\textbf{Bayesian methods}: MC Dropout obtains an RMSE improvement for selective prediction, resulting in 4.11 kts at the cost of a coverage of approximately 0 \%. This means that after selective predictions almost no samples remain, potentially adapting the threshold may result in improvements. SWAG obtains significant improvements with selective prediction at the cost of a low coverage of 11\% and obtains relatively low CRPS and NLL as well as MACE averaged over categories, see Table \ref{tab:results}, as well as per category, Figure \ref{fig:uq_metrics_result}. This indicates a good fit, however the coverage after selective prediction may be improved with a different threshold. Laplace obtains no improvement with selective prediction and interestingly also has a constant sharpness across categories as Figure \ref{fig:uq_metrics_result} shows. This may be due to the fact that the Laplace approximation uses a second order Taylor expansion with respect to the model parameters of the loss and does not take into account variances in the data to construct a Gaussian approximation to the posterior weight distribution. BNN VI ELBO interestingly obtains the lowest RMSE per category and overall, Table 1 in the Appendix, Section 1, which indicates a good fit of the mean prediction. However, the predictive uncertainties are relatively small as the low sharpness and high negative log likelihood per category suggest, Figure \ref{fig:uq_metrics_result}. DKL obtains a relatively high RMSE and no improvement with selective prediction, although the correlation between predictive uncertainty and MAE,  Figure \ref{fig:uq_metrics_result}, is also high on higher categories. However this may be due to high errors and high uncertainties. Compared to DKL, DUE obtains a significantly lower RMSE which may be due to the spectral normalization of layers, as this is the only difference between the methods. Otherwise DUE obtains a lower MACE per category than DKL, yet also a lower correlation between predictive uncertainty and MAE.
\textbf{Diffusion UQ methods}, surprisingly CARD obtains a average RMSE and a significant improvement with selective prediction, while maintaining a coverage of 60 \% and a low miscalibration error (MACE) of 0.05.

\vspace{-5pt}
\section{Conclusion}\label{conclusion}
\vspace{-5pt}
We presented a first analysis of predictive uncertainty for cyclone wind speed estimation. The various methods considered performed quite differently across storm categories and often exhibited a tradeoff between coverage vs accuracy. When predicting the maximum sustained wind speed, MVE demonstrated high coverage and low RMSE.  Yet if a lower coverage is tolerable, then SWAG is a more attractive option due to it having a better RMSE than MVE.  In future work, we plan to consider autoregressive models for the time series task presented in Figure 3.

\section{Acknowledgements}\label{acknowledgement}
This work was supported by the Helmholtz Association's Initiative and Networking Fund on the HAICORE@KIT partition.

\bibliography{iclr2024_conference.bib}
\bibliographystyle{iclr2024_conference}

\end{document}

\begin{centering}\maketitle
\end{centering}

\section{Additional Figures and Tables}

\subsection{Experiments with a minimum wind speed of zero}

\begin{figure}[h!]
    \centering
    \includegraphics[width=14cm,height=7cm,keepaspectratio]{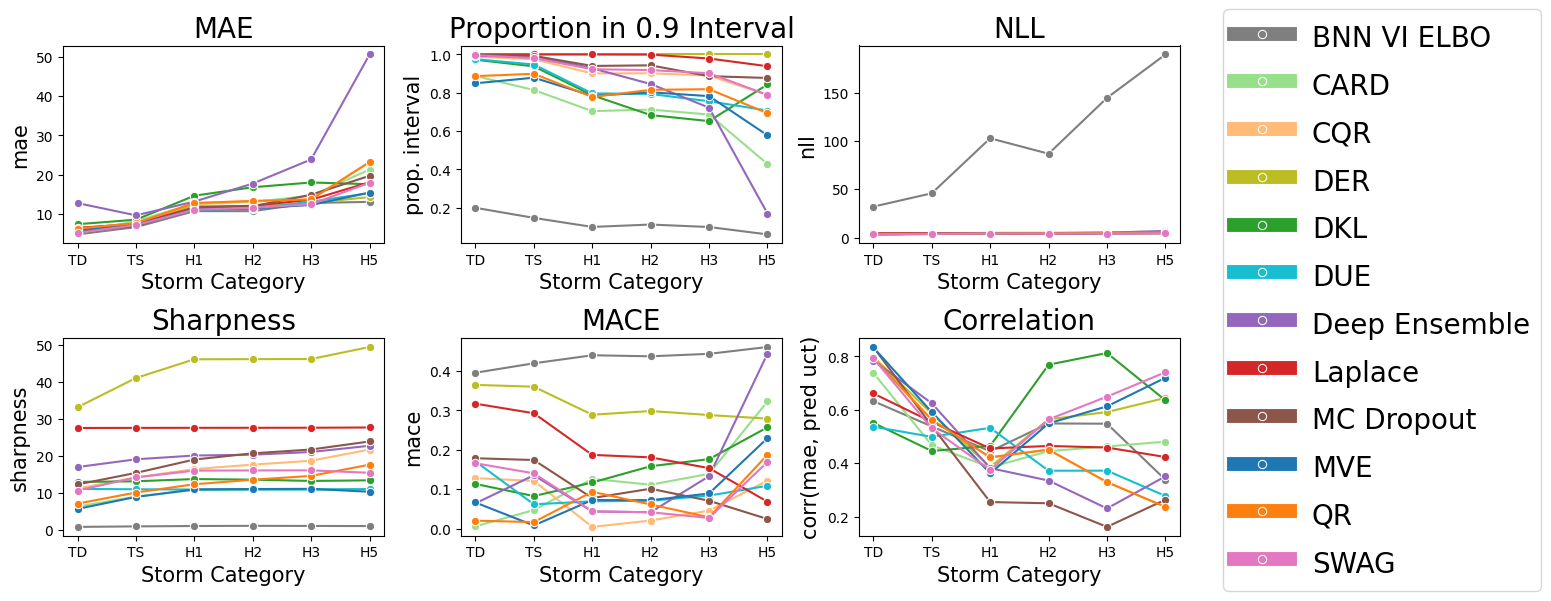}
    \caption{Uncertainty Metrics over different storm categories.}
    \label{fig:uq_metrics_result_all_cats}
\end{figure}

\begin{figure}[h!]
    \centering
    \includegraphics[width=14cm,height=7cm,keepaspectratio]{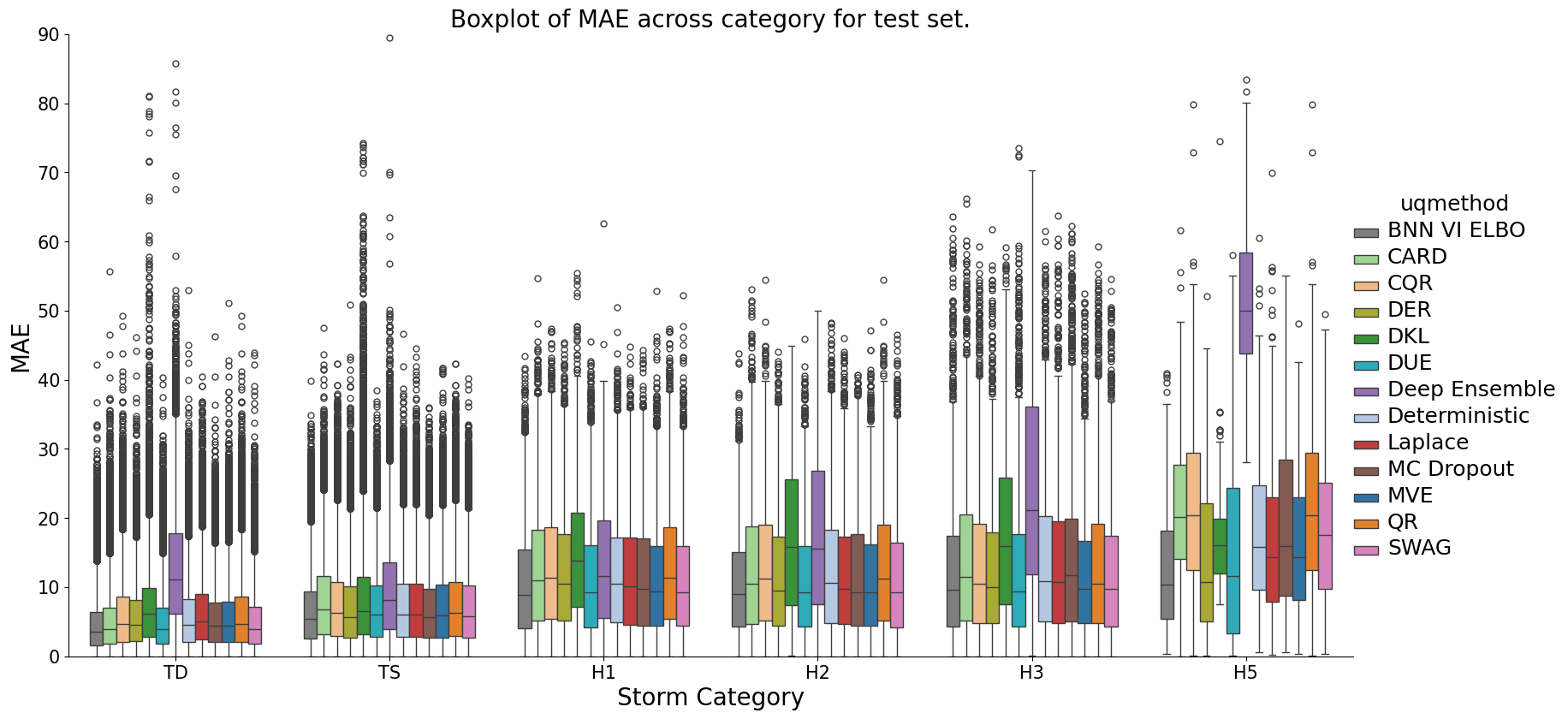}
    \caption{MAE over different storm categories.}
    \label{fig:mae_all_cats}
\end{figure}

\begin{figure}[h!]
    \centering
    \includegraphics[width=14cm,height=7cm,keepaspectratio]{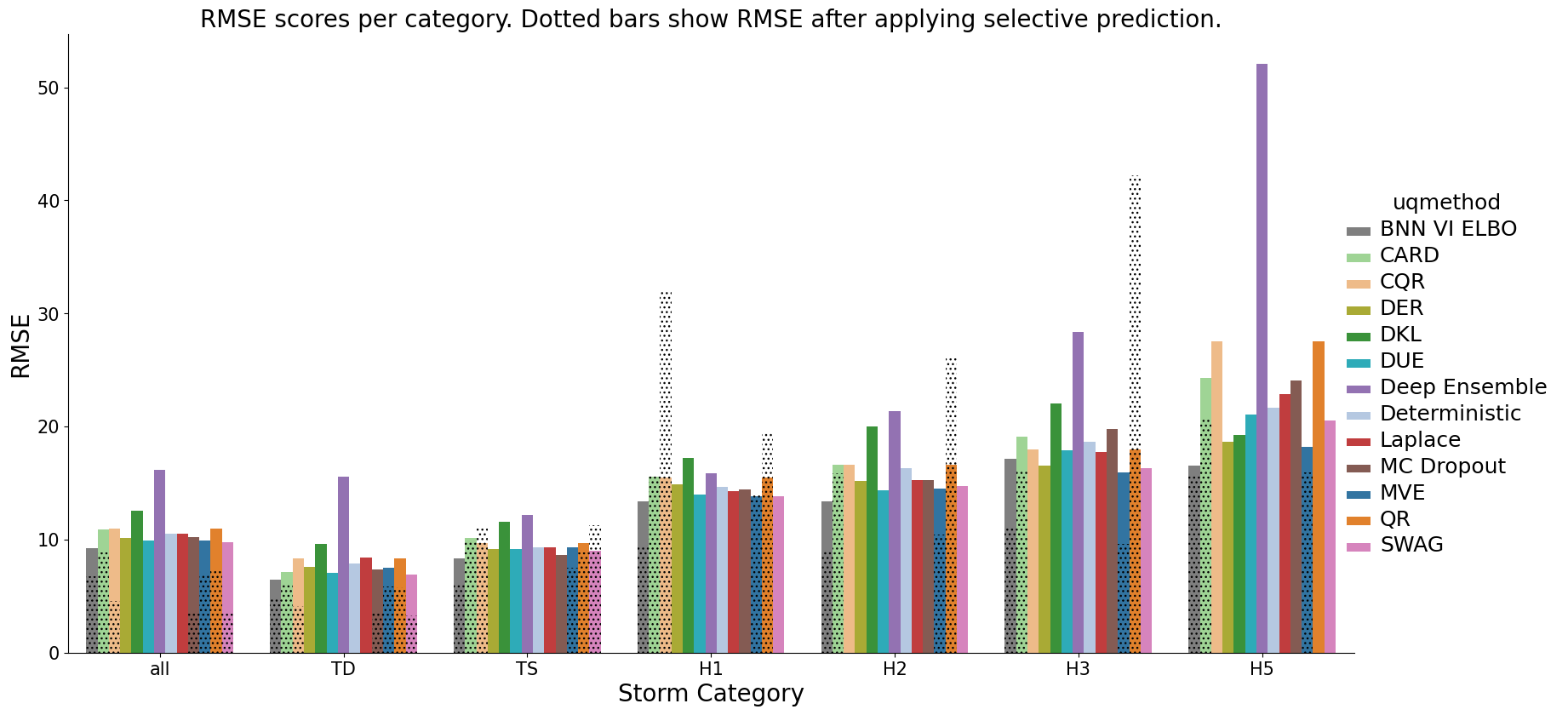}
    \caption{RMSE before and after selective prediction over different storm categories. The dotted bars are the RMSE after selective prediction with a threshold of 9 kts.}
    \label{fig:mae_all_cats}
\end{figure}

\begin{table}[H]
\centering
\resizebox{13cm}{!}{%
\begin{tabular}{ll|rr|rr|rr|rr|rr|rr|rr}
\toprule
 & & \multicolumn{2}{c}{TD} & \multicolumn{2}{c}{TS} & \multicolumn{2}{c}{H1} & \multicolumn{2}{c}{H2} & \multicolumn{2}{c}{H3} & \multicolumn{2}{c}{H4} & \multicolumn{2}{c}{H5} \\
& uqmethod  & rmse $\downarrow$ & nll $\downarrow$ & rmse $\downarrow$ & nll $\downarrow$ & rmse $\downarrow$ & nll $\downarrow$ & rmse $\downarrow$ & nll $\downarrow$ & rmse $\downarrow$ & nll $\downarrow$ & rmse $\downarrow$ & nll $\downarrow$ & rmse $\downarrow$ & nll $\downarrow$ \\
\midrule
\multirow{2}{*}{Deterministic} & MVE & 7.504 & \textbf{3.264} & 9.313 & 3.692 & 13.801 & 4.330 & 14.504 & 4.219 & \textbf{15.911} & 4.337 & \textbf{14.843} & \textbf{4.149} & 18.175 & 4.691 \\
& DER & 7.581 & 4.432 & 9.185 & 4.651 & 14.902 & 4.802 & 15.178 & 4.800 & 16.553 & 4.809 & 14.946 & 4.812 & 18.644 & 4.885 \\
\midrule
\multirow{2}{*}{Quantile} & CQR & 8.319 & 3.529 & 9.723 & 3.801 & 15.508 & 4.209 & 16.635 & 4.280 & 17.972 & 4.430 & 17.401 & 4.345 & 27.543 & 4.934 \\
& QR & 8.319 & 3.359 & 9.723 & 3.731 & 15.508 & 4.370 & 16.635 & 4.422 & 17.972 & 4.690 & 17.401 & 4.404 & 27.543 & 5.286 \\
\midrule
\multirow{1}{*}{Ensemble} &Deep Ensemble & 15.562 & 4.038 & 12.150 & 4.028 & 15.901 & 4.297 & 21.370 & 4.582 & 28.335 & 5.038 & 36.286 & 5.432 & 52.112 & 6.817 \\
\midrule
\multirow{6}{*}{Bayesian}
& \textbf{BNN VI ELBO} & \textbf{6.479} & 31.978 & \textbf{8.361} & 45.774 & \textbf{13.405} & 102.664 & \textbf{13.377} & 86.707 & 17.175 & 144.454 & 15.703 & 105.009 & \textbf{16.581} & 189.350 \\
& BNN VI & 8.152 & 3.422 & 8.918 & \textbf{3.617} & 14.409 & 4.355 & 17.683 & 4.613 & 22.457 & 5.288 & 25.110 & 5.004 & 33.325 & 5.224 \\
& Laplace & 8.401 & 4.281 & 9.343 & 4.293 & 14.308 & 4.371 & 15.299 & 4.390 & 17.773 & 4.444 & 16.493 & 4.416 & 22.843 & 4.580 \\
& MC Dropout & 7.374 & 3.565 & 8.643 & 3.808 & 14.401 & 4.204 & 15.263 & 4.256 & 19.763 & 4.501 & 19.454 & 4.475 & 24.063 & 4.624 \\
& DKL & 9.634 & 3.757 & 11.552 & 3.892 & 17.243 & 4.367 & 19.994 & 4.542 & 22.006 & 4.776 & 15.287 & 4.172 & 19.217 & 4.535 \\
& DUE & 7.036 & 3.523 & 9.139 & 3.662 & 13.948 & 4.126 & 14.375 & 4.177 & 17.870 & 4.651 & 16.312 & 4.427 & 21.046 & 5.177 \\
& \textbf{SWAG} & 6.940 & 3.427 & 9.051 & 3.775 & 13.805 & \textbf{4.103} & 14.761 & \textbf{4.115} & 16.307 & \textbf{4.187} & 16.083 & 4.155 & 20.497 & \textbf{4.495} \\
\midrule
\multirow{1}{*}{Diffusion} & CARD & 7.167 & 3.268 & 10.158 & 4.045 & 15.557 & 4.816 & 16.622 & 4.858 & 19.077 & 5.329 & 17.497 & 4.785 & 24.322 & 6.378 \\
\bottomrule
\end{tabular}}
\caption{Evaluation Results on test set. RMSE per category.}
\label{tab:results_catknts}
\end{table}

\newpage

\subsection{Experiments with a minimum wind speed of 34}

\begin{figure}[h!]
    \centering
    \includegraphics[width=14cm,height=7cm,keepaspectratio]{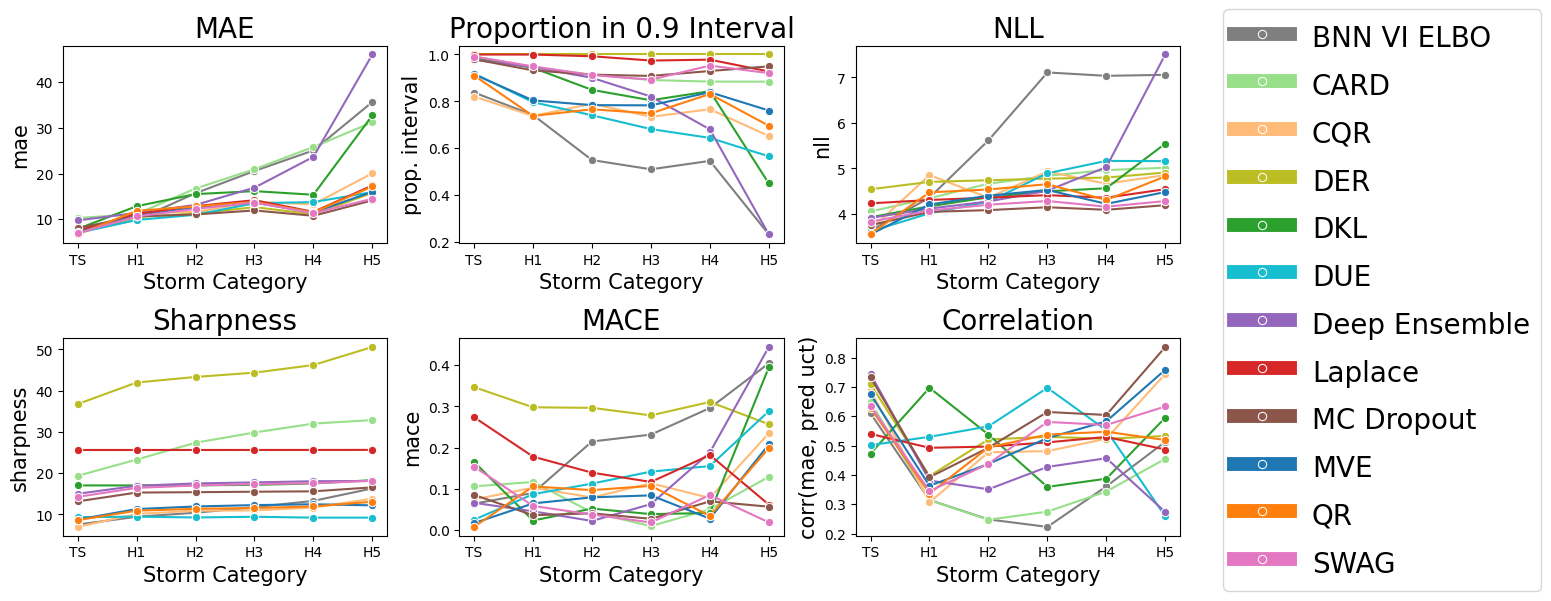}
    \caption{Uncertainty Metrics over different storm categories.}
    \label{fig:uq_metrics_result_all_cats}
\end{figure}

\begin{table}[H]
\centering
\resizebox{13cm}{!}{%
\begin{tabular}{llrrrrrr}
\toprule
UQ group & Method & RMSE $\downarrow$ & RMSE $\Delta$ $\uparrow$ & Coverage $\uparrow$ & CRPS $\downarrow$ & NLL $\downarrow$ & MACE $\downarrow$ \\
\midrule
None & Deterministic & 11.68 & 0.00 & \textbf{1.0}0 & NaN & NaN & NaN \\
\midrule
\multirow{2}{*}{Deterministic} & MVE & 11.21 & NaN & NaN & 6.08 & \textbf{3.79} & 0.03 \\
 & DER & \textbf{1\textbf{1.0}5} & NaN & 0.00 & 10.24 & \textbf{4.6}0 & 0.33 \\
\midrule
\multirow{2}{*}{Quantile} & QR & 11.33 & 2.26 & 0.44 & 6.11 & 3.86 & 0.03 \\
 & CQR & 11.57 & 1.99 & 0.62 & 6.25 & 4.00 & 0.08 \\
\midrule
\multirow{1}{*}{Ensemble} & Deep Ensemble & 14.56 & NaN & 0.00 & 7.96 & 4.06 & 0.04 \\
\midrule
\multirow{6}{*}{Bayesian} & MC Dropout & 11.56 & 2.55 & 0.01 & 6.40 & 3.85 & 0.07 \\
 & SWAG & 1\textbf{1.0}9 & NaN & NaN & 6.26 & 3.93 & 0.12 \\
 & Laplace & 11.68 & NaN & 0.00 & 7.98 & 4.27 & 0.24 \\
 & BNN VI ELBO & 13.10 & 1.73 & 0.68 & 7.11 & 4.20 & 0.10 \\
 & DKL & 13.17 & NaN & NaN & 7.43 & 4.05 & 0.11 \\
 & DUE & 1\textbf{1.0}7 & NaN & 0.00 & \textbf{6.03} & 3.85 & \textbf{0.02} \\
\midrule
\multirow{1}{*}{Diffusion} & CARD & 15.09 & \textbf{4.6}0 & \textbf{0.02} & 8.83 & 4.22 & 0.09 \\
\bottomrule
\end{tabular}%
}
\caption{Evaluation Results on test set. RMSE $\Delta$ shows the improvement after selective prediction, while Coverage denotes the fraction of remaining samples that were not omitted. Threshold 9 knots.}
\label{tab:results_934knts}
\end{table}

\newpage

\subsection{Experiments with a minimum wind speed of 64}

\begin{figure}[h!]
    \centering
    \includegraphics[width=14cm,height=7cm,keepaspectratio]{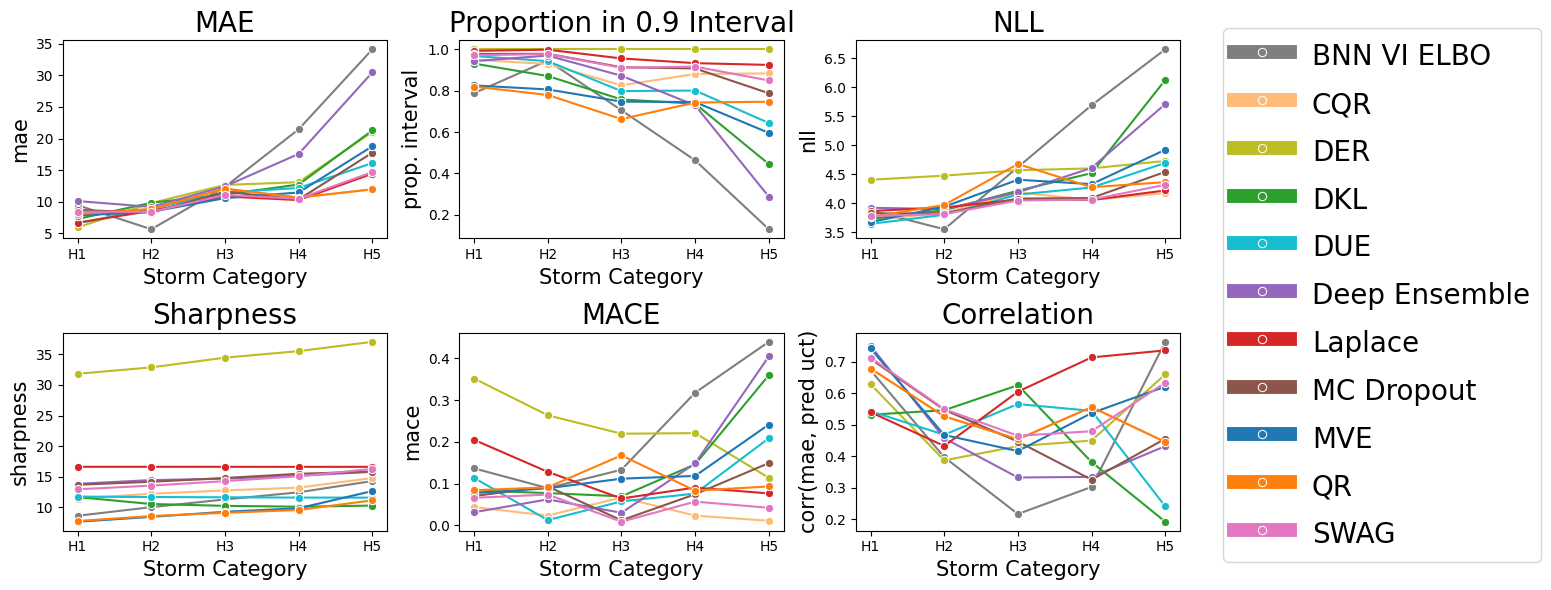}
    \caption{Uncertainty Metrics over different storm categories.}
    \label{fig:uq_metrics_result_all_cats}
\end{figure}

\begin{table}[H]
\centering
\resizebox{13cm}{!}{%
\begin{tabular}{llrrrrrr}
\toprule
UQ group & Method & RMSE $\downarrow$ & RMSE $\Delta$ $\uparrow$ & Coverage $\uparrow$ & CRPS $\downarrow$ & NLL $\downarrow$ & MACE $\downarrow$ \\
\midrule
None & Deterministic & 10.78 & 0.00 & \textbf{1.0}0 & NaN & NaN & NaN \\
\midrule
\multirow{2}{*}{Deterministic} & MVE & 11.48 & 1.43 & 0.64 & 6.45 & 3.95 & 0.09 \\
 & DER & 11.90 & NaN & 0.00 & 9.33 & 4.47 & 0.29 \\
\midrule
\multirow{2}{*}{Quantile} & QR & 11.76 & 1.20 & 0.62 & 6.68 & 4.03 & 0.10 \\
 & CQR & 11.76 & \textbf{2.46} & 0.05 & 6.55 & 3.87 & 0.02 \\
\midrule
\multirow{1}{*}{Ensemble} & Deep Ensemble & 14.43 & NaN & 0.00 & 8.08 & 4.07 & \textbf{0.01} \\
\midrule
\multirow{6}{*}{Bayesian} & MC Dropout & 11.86 & NaN & 0.00 & 6.70 & 3.91 & 0.06 \\
 & SWAG & 11.59 & NaN & 0.00 & 6.50 & 3.87 & 0.05 \\
 & Laplace & \textbf{10.74} & NaN & 0.00 & 6.37 & 3.94 & 0.15 \\
 & BNN VI ELBO & 13.82 & 2.44 & 0.44 & 7.87 & 4.19 & 0.11 \\
 & DKL & 12.12 & NaN & 0.00 & 6.80 & 3.98 & 0.02 \\
 & DUE & 11.38 & NaN & 0.00 & \textbf{6.31} & \textbf{3.85} & 0.03 \\
\bottomrule
\end{tabular}%
}
\caption{Evaluation Results on test set. RMSE $\Delta$ shows the improvement after selective prediction, while Coverage denotes the fraction of remaining samples that were not omitted. Threshold 9 knots.}
\label{tab:results_964knts}
\end{table}

\newpage

\section{Overview of applied UQ Methods}

We consider regression problems in the following setting: Given an input $x^\star \in X$ the task is to predict a target $y^\star \in Y$. Notably, regression is distinguishable from classification as the targets are continuous and possibly infinite, as opposed to a fixed set of finite labels in classification. In our experiments we want to use a neural network to predict a unobserved test target $y^\star \in Y$ for a given unobserved test input $x^\star \in X$. Precisely, given the set of $n \in \mathbb{N}$ observed training input-target pairs from our dataset,
 \begin{equation}
     \mathcal{D}_{\mathrm{train}} = \{(x_i, y_i)\}_{i=1}^{n},
 \end{equation} 
 the task of the models or NNs is to predict a target $y^\star \in Y$ given an input $x^\star \in X$ such that the loss objective between the predictions and targets is minimized over all the training points. This is described in the following.\\

The model or NN can be regarded as a function $f_{\theta}$, parameterized by weights $\theta$, that maps inputs $x$ directly to targets $y \in Y$, 
 \begin{equation}
     f_{\theta}: X \rightarrow Y
 \end{equation}
 or to a probability distribution, 
 \begin{equation}
     f_{\theta}: X \rightarrow \mathcal{P}(Y)
 \end{equation}
such that  
\begin{equation}
    f_{\theta}(x^\star)=p_{\theta}(x^\star) \in \mathcal{P}(Y).
\end{equation}
or as $f_{\theta}: X \rightarrow Y^n$ that maps to $n$ quantiles, \begin{equation}
    f_{\theta}(x^\star) = (q_1(x^\star),...,q_n(x^\star)) \in Y^n.
\end{equation}

For example, a NN can be configured to output the mean and standard deviation of a Gaussian distribution $f_{\theta}(x^\star)=(\mu_{\theta}(x^\star),\sigma_{\theta}(x^\star))$. \\

\begin{table}[h!]
\centering
\resizebox{\columnwidth}{!}{%
\begin{tabular}{ l|c|c|c|c|c  }
 \multicolumn{6}{c}{Previous Benchmarks and Reviews of Uncertainty Quantification Methods for Regression Problems} \\
 \hline
Publication & \citep{gustafsson2023Reliable} &  \citep{schmahling2022framework} & \citep{dewolf2022valid} & \citep{izmailov2021bayesian}   & here \\
 \hline
\textbf{Deterministic Methods} &  &  &  & &  \\
 \hline
 Baseline & & & & &  \checkmark \\
 Gaussian (MVE) & \checkmark  &  & & &  \checkmark \\
 Deep Evidential Networks (DER) & & & & &  \checkmark \\
 \hline
\textbf{ Ensemble based } &  &  &  & &  \\
 \hline
Deep Ensembles, GMM  & \checkmark  & \checkmark  & \checkmark & &  \checkmark \\
\hline
\textbf{Bayesian}  &  &  &  & &  \\
\hline
MC Dropout, GMM & & \checkmark  &  \checkmark & &  \checkmark \\
BNN with VI &  &   & &  \checkmark & \checkmark \\
%BNN + LV & & & & &  \checkmark \\
Laplace Approximation & &  & & &  \checkmark \\
SWAG & &  & &  \checkmark& \checkmark \\
%Mulit-Swag &  &  & & &  \checkmark \\
%SGLD & & & & \checkmark & \checkmark \\
DVI, SI & &  & &  \checkmark& \\
HMC & &  & &  \checkmark & \\
\hline
\textbf{Gaussian Process based} &  &  &  & &  \\
\hline
"Gaussian Process (GP)" & &  & \checkmark & &  \\
Approximate GP & &  & \checkmark  & &  \\
Deep Kernel Learning (DKL) & & & & & \checkmark \\
Spectral Normalized GPs (DUE) & & & & &  \checkmark \\
\hline
\textbf{Quantile based} &  &  &  & &  \\
\hline
Quantile Regression (QR) & \checkmark  &  & \checkmark & & \checkmark \\
Conformal Prediction (CQR) & \checkmark  & & \checkmark & & \checkmark \\
 \hline
\textbf{Diffusion Model} &  &  &  & &  \\
\hline
CARD &  &  &  & & \checkmark \\
\end{tabular}%
}
\caption{Comparison of previous reviews. The BNN implementations of BNN with VI and SWAG in this work use partially stochastic networks, as proposed in \citep{sharma2023bayesian}.}
\label{table:comparisonreviews}
\end{table}

 As Table \ref{table:comparisonreviews} demonstrates, we compare five classes of UQ methods: deterministic, ensemble, Bayesian, quantile and diffusion based methods. Firstly, deterministic UQ methods use a DNN, $f_{\theta}: X \rightarrow \mathcal{P}(Y)$, that map inputs $x$ to the parameters of a probability distribution $f_{\theta}(x^\star)=p_{\theta}(x^\star) \in \mathcal{P}(Y)$. These include Deep Evidential Networks (\textbf{DER}) \cite{amini2020deep}, where we use the correction proposed by \cite{meinert2023unreasonable}, and Mean Variance Networks (\textbf{MVE}) \citep{nix1994estimating} which output the mean and standard deviation of a Gaussian distribution $f_{\theta}^{\text{MVE}}(x^\star)=(\mu_{\theta}(x^\star),\sigma_{\theta}(x^\star))$. Secondly, the broadly considered state-of-the-art method Deep Ensembles (\textbf{DeepEnsembles}) proposed by \cite{lakshminarayanan2017simple} utilizes an ensemble over MVE networks. Thirdly, Bayesian methods aim at modelling a distribution over the network parameters and are commonly used to approximate the first and second moment of a marginalized distribution. These include Bayesian Neural Networks with Variational Inference (\textbf{BNN VI ELBO}) \cite{blundell2015weight}, MC-Dropout (\textbf{MCDropout}) \cite{gal2016dropout}, the Laplace Approximation (\textbf{Laplace}) \cite{ritter2018scalable}\cite{daxberger2021laplace} and \textbf{SWAG} \cite{maddox2019simple}. A slightly different approach is taken by Gaussian process based methods that model a distribution over functions, yet also approximate the fist and second moment of this distribution. These include Deep Kernel Learning (\textbf{DKL}) \cite{wilson2016deep} and an extension thereof Deterministic Uncertainty Estimation (\textbf{DUE}) \citep{van2021feature}. Fourthly, quantile based models $f_{\theta}: X \rightarrow Y^n$ that map to $n$ quantiles, $f_{\theta}(x^\star) = (q_1(x^\star),...,q_n(x^\star)) \in Y^n$, such as Quantile Regression (\textbf{Quantile Regression}) and the conformalized version thereof (\textbf{ConformalQR}) suggested by \cite{romano2019conformalized}. Lastly, we also consider a diffusion model (\textbf{CARD}) as introduced by \cite{han2022card}. A detailed description of the methods is provided in the supplementary material.

\section{Description of UQ Methods}

\bigskip
 \textbf{Baseline model}: Depending on the application a DNN, for example a residual network, that is used as a baseline. This model does not predict any uncertainty and just a mean $f_{\theta}(x^{\star})$. For the loss objective, the mean squared error is used
 
 \begin{equation}
     \mathcal{L}(\theta, (x^{\star}, y^{\star})) = (f_{\theta}(x^{\star}) -y^{\star})^2.
 \end{equation}

\subsection{Deterministic UQ Methods}

In the following we list the deterministic UQ methods considered in this work. These methods provide UQ estimates within a single forward pass by predicting the parameters of a probability distribution.

\bigskip
\textbf{Gaussian}: The Gaussian model, also referred to as Mean Variance Estimation, first studied in \cite{nix1994estimating} and further used in \cite{sluijterman2023optimal}, is a deterministic model that predicts the parameters of a Gaussian distribution 

\begin{equation}
    f_{\theta}(x^{\star}) = (\mu_{\theta}(x^\star),\sigma_{\theta}(x^\star))
\end{equation}

in a single forward pass, where standard deviations $\sigma_{\theta}(x^\star)$ can be used as a measure of data uncertainty. To this end, the network now outputs two parameters and is trained with the Gaussian negative log-likelihood (NLL) as a loss objective \cite{kendall2017uncertainties}, that is given by

\begin{equation}
    \mathcal{L}(\theta, (x^{\star}, y^{\star})) = \frac{1}{2}\text{ln}\left(2\pi\sigma_{\theta}(x^{\star})^2\right) + \frac{1}{2\sigma_{\theta}(x^{\star})^2}\left(\mu_{\theta}(x^{\star})-y^{\star}\right)^2.
\end{equation}  

Correspondingly, the model prediction consists of a predictive mean, $\mu_{\theta}(x^\star)$, and the predictive uncertainty, in this case the standard deviation $\sigma_{\theta}(x^\star)$.

\bigskip
\textbf{Deep Evidential Networks (DER)}:

Deep Evidential Regression (DER) \citep{amini2020deep} is a single forward pass UQ method that aims to disentangle aleatoric and epistemic uncertainty.  DER entails a four headed network output \begin{equation}
    f_{\theta}(x^{\star})=(\gamma_{\theta}(x^{\star}), \nu_{\theta}(x^{\star}), \alpha_{\theta}(x^{\star}), \beta_{\theta}(x^{\star})),
\end{equation}

that is used to compute the predictive t-distribution with $2\alpha(x^{\star})$ degrees of freedom:

\begin{equation}\label{eq:evi}
    p(y(x^{\star})|f_{\theta}(x^{\star}))=\text{St}_{2\alpha_{\theta}(x^{\star})}\left(y^{\star}\bigg| \gamma_{\theta}(x^{\star}), \frac{\beta_{\theta}(x^{\star})(1+\nu_{\theta}(x^{\star}))}{\nu_{\theta}(x^{\star})\alpha_{\theta}(x^{\star})}\right).
\end{equation}

In \cite{amini2020deep} the network weights are obtained by minimizing the loss objective that is the negative log-likelihood of the predictive distribution and a regularization term. However, due to several drawbacks of DER, \cite{meinert2023unreasonable} propose the following adapted loss objective that we also utilise,

\begin{equation}
    \mathcal{L}({\theta},(x^{\star}, y^{\star}))=\log\sigma_{\theta}^2(x^{\star})+(1+\lambda\nu_{\theta}(x^{\star}))\frac{(y^{\star}-\gamma_{\theta}(x^{\star}))^2}{\sigma_{\theta}^2(x^{\star})}
\end{equation}
where $\sigma_{\theta}^2(x^{\star})=\beta_{\theta}(x^{\star})/\nu_{\theta}(x^{\star})$. The mean prediction is given as,

\begin{equation}
    \mu_{\theta}(x^{\star}) = \gamma_{\theta}(x^{\star}).
\end{equation}

Further following \citep{meinert2023unreasonable}, we use their reformulation of the uncertainty decomposition. The aleatoric uncertainty is given by

\begin{equation}
    u_{\text{aleatoric}}(x^{\star})=\sqrt{\frac{\beta(x^{\star})}{\alpha(x^{\star})-1}},
\end{equation}

and the epistemic uncertainy by,

\begin{equation}
    u_{\text{epistemic}}(x^{\star})=\frac{1}{\sqrt{\nu(x^{\star})}}.
\end{equation}

The predictive uncertainty is then, given by 

\begin{equation}
    u(x^{\star}) = \sqrt{u_{\text{epistemic}}(x^{\star})^2+u_{\text{aleatoric}}(x^{\star})^2} .
\end{equation}

\subsection{Ensemble Based UQ Methods}

\bigskip
\textbf{Deep Ensembles}: 
introduced in \cite{lakshminarayanan2017simple}, Deep Ensembles approximate a posterior distribution over the model weights with a Gaussian mixture model over the output of separately initialized and trained networks. In \cite{wilson2020bayesian} the authors showed that Deep Ensembles can be interpreted as a Bayesian method. \\

For the Deep Ensembles model the predictive mean is given by the mean taken over $N \in \mathbb{N}$ models $f_{\theta_i}(x^{\star}) = \mu_{\theta_i}(x^{\star})$ that output a mean with different weights $\{\theta_i\}_{i=1}^N$, 

\begin{equation}
     \mu(x^{\star}) = \frac{1}{N} \sum_{i=1}^N  \mu_{\theta_i}(x^{\star}).
\end{equation}

The predictive uncertainty is given by the standard deviation of the predictions of the $N$ different networks, Gaussian ensemble members,

\begin{equation}
    \sigma(x^{\star}) = \sqrt{\frac{1}{N} \sum_{i=1}^N  \left(\mu_{\theta_i}(x^{\star})-  \mu(x^{\star}) \right)^2}.
\end{equation}

\begin{table}[h!]
\centering
\resizebox{\columnwidth}{!}{%
\begin{tabular}{ l|c|l }
 \multicolumn{3}{c}{Summary of hyperparameters for the Deep Ensembles model} \\
 \hline
Hyperparameter & value range & hints  \\
 \hline
Number of ensemble members & $N \approx [5,20]$& do an ablation study on $N$. \\
 \hline
\end{tabular}%
}
\label{table:hypers4}
\end{table}

\bigskip
\textbf{Deep Ensembles GMM}:

For the Deep Ensembles GMM model, the predictive mean is given by the mean taken over $N \in \mathbb{N}$ models $f_{\theta_i}(x^{\star}) = (\mu_{\theta_i}(x^{\star}), \sigma_{\theta_i}(x^{\star}))$ with different weights $\{\theta_i\}_{i=1}^N$, 

\begin{equation}
     \mu_g(x^{\star}) = \frac{1}{N} \sum_{i=1}^N  \mu_{\theta_i}(x^{\star}).
\end{equation}

The predictive uncertainty is given by the standard deviation of the Gaussian mixture model consisting of the $N$ different networks, Gaussian ensemble members,

\begin{equation}
    \sigma_g(x^{\star}) = \sqrt{\frac{1}{N} \sum_{i=1}^N  \left(\mu_{\theta_i}(x^{\star})-  \mu_g(x^{\star}) \right)^2+ \frac{1}{N}  \sum_{i=1}^N \sigma_{\theta_i}^2(x^\star)}.
\end{equation}

Note that the difference between "Deep Ensembles" and "Deep Ensembles GMM" is that in the latter we also consider the predictive uncertainty output of each individual ensemble member, whereas in the former we only consider the means and the variance of the mean predictions of the ensemble members.

Because each ensemble member has a probabilistic predictive distribution 
$(\mu_{\theta_i}(x^{\star}), \sigma_{\theta_i}(x^{\star}))$, we can also perform a decomposition into epistemic and aleatoric components:
\begin{align}
    u_{\text{epistemic}}(x^{\star})  &= \frac{1}{N} \sum_{i=1}^N  (\mu_g(x^{\star}) -  \mu_{\theta_i}(x^{\star}))^2 \;, \\
   u_{\text{aleatoric}}(x^{\star})  & = \frac{1}{N} \sum_{i=1}^N  \sigma_{\theta_i}^2(x^\star)  \;.
\end{align}

\begin{table}[h!]
\centering
\resizebox{\columnwidth}{!}{%
\begin{tabular}{ l|c|l }
 \multicolumn{3}{c}{Summary of hyperparameters for the Deep Ensembles model} \\
 \hline
Hyperparameter & value range & hints  \\
 \hline
Number of ensemble members & $N \approx [5,20]$& do an ablation study on $N$. \\
 \hline
\end{tabular}%
}
\label{table:hypers5}
\end{table}

\subsection{Bayesian UQ Methods}

The general aim of Bayesian UQ methods is to obtain the predictive distribution by marginalization over the model weights $\theta$,

\begin{equation}\label{eq:pred}
    p(y^{\star}|x^{\star},D) = \int p(y^{\star}|x^{\star}, \theta) p(\theta|D) d\theta.
\end{equation}

The posterior distribution over the weights $p(\theta|D)$ can be approximated by utilizing Bayes' theorem or, for example, by a variational approach. However, the predictive distribution, \eqref{eq:pred}, is usually intractable and, in the following various approaches of approximation are presented (most of which rely on sampling over the posterior).

\bigskip
\textbf{MC-Dropout}: Is an approximate Bayesian method with sampling. A fixed dropout rate $p \in [0,1)$ is used, meaning that random weights are set to zero during each forward pass with the probability $p$. This models the network weights and biases as a Bernoulli distribution with dropout probability $p$. While commonly used as a regularization method, \cite{gal2016dropout} showed that activating dropout during inference over multiple forward passes yields an approximation to the posterior over the network weights. Due to its simplicity it is widely adopted in practical applications, but MC-Dropout and variants thereof have also been criticized for their theoretical shortcomings \cite{hron2017variational}, \cite{osband2016risk}.\newline

For the MC Dropout model the prediction consists of a predictive mean and a predictive uncertainty. For the predictive mean, the mean is taken over $m \in \mathbb{N}$ forward passes through the network $f_{p,\theta}$ with a fixed dropout rate $p$, resulting in different weights $\{\theta_i\}_{i=1}^m$, given by

\begin{equation}
     f_p(x^{\star}) = \frac{1}{m} \sum_{i=1}^m  f_{p,\theta_i}(x^{\star}).
\end{equation}

The predictive uncertainty is given by the standard deviation of the predictions over $m$ forward passes,

\begin{equation}
    \sigma_p(x^{\star}) = \sqrt{\frac{1}{m} \sum_{i=1}^m  \left(f_{p,\theta_i}(x^{\star})-  f_p(x^{\star}) \right)^2}.
\end{equation}

\begin{table}[h!]
\centering
%\resizebox{\columnwidth}{!}{%
\begin{tabular}{ l|c|l }
 \multicolumn{3}{c}{Summary of hyperparameters for the MC Dropout model} \\
 \hline
Hyperparameter & value range & hints  \\
 \hline
Drop out rate & $p \in [0,1)$ & start with $p = 0.2$.\\
 \hline
\end{tabular}%
%}
\label{table:hypers6}
\end{table}

\bigskip
\textbf{MC Dropout GMM}:
We also consider combining this method with the previous model Gaussian network, as in \cite{kendall2017uncertainties}, aiming at disentangling the data and model uncertainties, abbreviated as MC Dropout GMM. For the MC Dropout GMM model,  the prediction again consists of a predictive mean and a predictive uncertainty $f_{p,\theta}(x^{\star}) = (\mu_{p,\theta}(x^{\star}), \sigma_{p,\theta}(x^{\star}))$. Here the predictive mean is given by the mean taken over $m$ forward passes through the Gaussian network mean predictions $\mu_{p,\theta}$ with a fixed dropout rate $p$, resulting in different weights $\{\theta_i\}_{i=1}^m$, given by

\begin{equation}
     \mu_p(x^{\star}) = \frac{1}{m} \sum_{i=1}^m  \mu_{p,\theta_i}(x^{\star}).
\end{equation}

The predictive uncertainty is given by the standard deviation of the Gaussian mixture model obtained by the predictions over $m$ forward passes,

\begin{equation}
    \sigma_p(x^{\star}) = \sqrt{\frac{1}{m} \sum_{i=1}^m  \left(\mu_{p,\theta_i}(x^{\star})-  \mu_p(x^{\star}) \right)^2 + \frac{1}{m}  \sum_{i=1}^m \sigma_{p,\theta_i}^2(x^\star)}.
\end{equation}

A decomposition of uncertainty can then be performend in a similar way as to with deep ensembles.

\begin{table}[h!]
\centering
%\resizebox{\columnwidth}{!}{%
\begin{tabular}{ l|c|l }
 \multicolumn{3}{c}{Summary of hyperparameters for the MC Dropout GMM model} \\
 \hline
Hyperparameter & value range & hints  \\
 \hline
Number of burn-in-epochs & $\approx[0, n]$ & after burn-in-epochs train variance and mean outputs. \\
Drop out rate & $p \in [0,1)$ & start with $p = 0.2$.\\
 \hline
\end{tabular}%
%}
\label{table:hypers7}
\end{table}

\bigskip
\textbf{BNN with VI}:
 Bayesian Neural Networks (BNNs) with variational inference (VI) are an approximate Bayesian method. Here, we follow the  mean-field assumption,  meaning that the variational distribution is factorized as a product of individual Gaussian distributions. This results in a diagonal Gaussian approximation of the posterior distribution over the model parameters
 
 The most common approach is to  maximize the evidence lower bound (ELBO). We note that there are other, alternative approaches for variational inference, such as $\alpha$-divergence minimization \citep{hernandez2016black}. 
 
 Uitilizing standard stochastic gradient descent by using the reparameterization trick \cite{kingma2013auto} one can  backpropagate through the necessary sampling procedure, a process called  Monte Carlo variational Bayes \citep{ranganath2014black}. \newline
 
The predictive likelihood is given by,

\begin{equation}
p(Y |\theta ,X) =
\prod_{i=1}^N p(y_i|\theta,x_i) =
\prod_{i=1}^N \mathcal{N}(y_{i} | f_{\theta}(x_i), \Sigma).
\end{equation}

The prior on the weights is given by,

\begin{equation}\label{eq:weightdist}
    p(\theta) = \prod_{l=1}^L \prod_{h=1}^{V_l }\prod_{j=1}^{V_{l-1}+1} \mathcal{N}(w_{hj, l} \vert 0, \lambda)
\end{equation}
where $w_{hj, l}$ is the h-th row and the j-th column of weight matrix $\theta_L$ at layer index $L$ and $\lambda$ is the prior variance. Note that as we use partially stochastic networks, \eqref{eq:weightdist} may contain less factors $\mathcal{N}(w_{hj, l} \vert 0, \lambda)$ depending on how many layers are stochastic. Then, the posterior distribution of the weights is obtained by Bayes' rule as

\begin{equation}\label{eq:posttrue}
    p(\theta|\mathcal{D}) = \frac{p(Y |\theta ,X) p(\theta)}{p(Y | X)}.
\end{equation}

As the posterior distribution over the weights is intractable a variational approximation is used,

\begin{equation}\label{eq:postweights}
    q(\theta) \approx p(\theta|\mathcal{D}),
\end{equation}

that is a diagonal Gaussian. Now given an input $x^{\star}$, the predictive distribution can be obtained as

\begin{equation}\label{eq:posteasy}
    p(y^{\star}|x^{\star},\mathcal{D}) = \int p(y^{\star} |\theta , x^{\star})  p(\theta|\mathcal{D}) d\theta.
\end{equation}

As \eqref{eq:posteasy} is intractable it is approximated by sampling form the approximation $q(\theta)$ in \eqref{eq:postweights} to the posterior distribution in \eqref{eq:posttrue}. The parameters of $q(\theta)$ are obtained by maximizing the evidence lower bound (ELBO) on the Kullback-Leibler (KL) divergence between the variational approximation and the posterior distribution over the weights. The negative ELBO is given by,

\begin{equation}
    \mathcal{L}(\theta, (x^{star}, y^{star}) ) = \beta D_{KL}(q(\theta) || p(\theta) ) + \frac{1}{2}\text{ln}\left(2\pi\sigma^2\right) + \frac{1}{2\sigma^2}\left(f_{\theta}(x^{\star})-y^{\star}\right)^2 . \label{eq:elbo}
\end{equation} 

The KL divergence can be computed analytically in the case of a Gaussian prior and the hyperparameter $\beta$ can be used to weight the influence of the variational parameters relative to that of the data. Alternatively, in the case of a fixed dataset of size $N$ this parameter is automatically set to $\frac{1}{N}$.
The hyperparameter $\sigma$ can be either fixed or set to be an additional parameter to be tuned by including it in the objective function Eq. (\ref{eq:elbo}), a process called type-II maximum likelihood.

The predictive mean is obtained as the mean of the network output $f_{\theta}$ with $S$ weight samples from the variational approximation $\theta_s \sim q(\theta)$,

\begin{equation}
     f_m(x^{\star}) = \frac{1}{S} \sum_{i=1}^S  f_{\theta_s}(x^{\star}).
\end{equation}

The predictive uncertainty is given by the standard deviation thereof, including the (possibly estimated) constant output noise $\sigma$:
\begin{equation}
    \sigma_p(x^{\star}) = \sqrt{\frac{1}{S} \sum_{i=1}^S  \left(f_{\theta_s}(x^{\star})-  f_m(x^{\star}) \right)^2 + \sigma^2}.
\end{equation}

If one uses the NLL and adapts the BNN to output a mean and standard deviation of a Gaussian $f_{\theta_s}(x^{\star}) = (\mu_{\theta_s}(x^{\star}), \sigma_{\theta_s}(x^{\star}))$, the mean prediction is given by

\begin{equation}
    f_m(x^{\star}) = \frac{1}{S} \sum_{s=1}^S  \mu_{\theta_s}(x^{\star}).
\end{equation}

and the predictive uncertainty is obtained as the standard deviation of the corresponding Gaussian mixture model obtained by the weight samples,

\begin{equation}
    \sigma_p(x^{\star}) = \sqrt{\frac{1}{S} \sum_{s=1}^S  \left(\mu_{\theta_s}(x^{\star})-  f_m(x^{\star}) \right)^2+ \sum_{s=1}^S \sigma_{\theta_s}^2(x^{\star})}.
\end{equation}

\begin{table}[h!]
\centering
\resizebox{\columnwidth}{!}{%
\begin{tabular}{ l|c|l }
 \multicolumn{3}{c}{Summary of hyperparameters for the BNN with VI model} \\
 \hline
Hyperparameter & value range & hints  \\
 \hline
Number burn-in-epochs & $\approx[0, n]$ & after burn-in-epochs train variance and mean outputs. \\
Loss scale factor $\beta$ & $\beta \approx [100,500]$ & should depend on parameter and train set size. \\
Samples during training $S_{tr}$ & $S_{tr} \approx [5, 20]$ & depending on network size and computing resources. \\
Samples during tests and prediction $S_{te}$ & $S_{te} \approx [5, 50]$ & depending on network size and computing resources. \\
Output noise scale $\sigma$ & $\sigma \approx [1.0, 5.0]$ & depending on label noise. \\
Prior mean $\mu_p$ for stochastic parameters & $\mu_p \approx [0,1.0]$ & start with $0$. 
Prior variance $\sigma_p$ for stochastic parameters & $\sigma_p \approx [0,3.0]$ & start with $1.0$. \\
Mean initialization for posterior $\mu_{pr}$ & $\mu_{pr} \approx [0, 1.0]$ & approximate posterior over parameters \\
Variance initialization for posterior $\rho_{pr}$ & $\rho_{pr} \approx [-6.0, 0.0]$  & variance through $\sigma = \log(1+\exp(\rho))$, approximate posterior over parameters\\
Bayesian layer type & "flipout" or "reparameterization" & \\
Stochastic module names & list of module names or a list of module numbers &  Transform module to be stochastic. \\
 \hline
\end{tabular}%
}
\label{table:hypers8}
\end{table}

\bigskip
\textbf{Laplace Approximation}: Originally introduced by \cite{mackay1992practical}, the Laplace Approximation has been adapted to modern neural networks by \cite{ritter2018scalable} and \cite{daxberger2021laplace} and is an approximate Bayesian method. The goal of the Laplace Approximation is to use a second-order Taylor expansion around the fitted MAP estimate and yield a posterior approximation over the model parameters via a full-rank, diagonal or Kronecker-factorized approach. In order for the Laplace Approximation to be computationally feasible for larger network architectures, we use the \href{https://aleximmer.github.io/Laplace/}{Laplace library} to include approaches, such as subnetwork selection that have been for example proposed by \cite{daxberger2021bayesian}.\newline

The general idea of the Laplace Approximation to obtain a distribution over the network parameters with a Gaussian distribution centered at the MAP estimate of the parameters \cite{daxberger2021bayesian}. In this setting, a prior distribution $p(\theta)$ is defined over our network parameters. Because modern neural networks consists of millions of parameters, obtaining a posterior distribution over the weights $\theta$ is intractable. The LA takes MAP estimate of the parameters $\theta_{MAP}$ from a trained network $f_{\theta_{MAP}}(x) = \mu_{\theta_{MAP}}(x)$ and constructs a Gaussian distribution around it. The parameters $\theta_{MAP}$ are obtained by
\begin{equation}
    \theta_{MAP} = \text{argmin} \mathcal{L}(\theta; D),
\end{equation} 

where $\mathcal{L}$ is the mean squared error or also referred to as the $\ell^2$ loss, $\mathcal{L}(\theta; \mathcal{D}) := -\sum_{i=1}^n log(p(y_i|f_{\theta}(x_i)))$ and the posterior $p(y_i|f_{\theta}(x_i))$ is chosen to be a Gaussian with constant variance $\sigma^2$, such that the loss is the mean squared error and a homoskedastic noise model is assumed. Then with Bayes Theorem, as in  \cite{daxberger2021bayesian}, one can relate the posterior to the loss,

\begin{equation}
    p(\theta|D) = p(D\vert\theta)p(\theta)/p(D)= \frac{1}{Z} exp(- \mathcal{L}(\theta; D)),
\end{equation}

with $Z = \int p(D\vert\theta)p(\theta) d\theta$. Now a second-order expansion of $\mathcal{L}$ around $\theta_{MAP}$ is used to construct a Gaussian approximation to the posterior $p(\theta|D)$:

\begin{equation}
    -\mathcal{L}(\theta; D) \approx -\mathcal{L}(\theta_{MAP}; D)- \frac{1}{2}(\theta-\theta_{MAP}) (\nabla_{\theta}^2 \mathcal{L}(\theta; D)\vert \theta_{MAP}) (\theta-\theta_{MAP}).
\end{equation}

The term with the first order derivative is zero as the loss is evaluated at a minimum $\theta_{MAP}$ \cite{murphy2022probabilistic}, and, further, one assumes that the first term is neglible as the loss is evaluated at $\theta = \theta_{MAP}$. Then taking the expontential of both sides allows to identify, after normalization, the Laplace approximation,

\begin{align}
p(\theta|D) \approx \mathcal{N}(\theta_{MAP}, \Sigma) && \text{with} \qquad \Sigma =  (\nabla_{\theta}^2 \mathcal{L}(\theta; D)\vert \theta_{MAP})^{-1}.
\end{align}

As the covariance is just the inverse Hessian of the loss, with $\theta_{MAP}\in \mathcal{R}^W$ and $H^{-1}\in \mathcal{R}^{W\times W}$, with $W$ being the number of weights, the posterior distribution is given by

\begin{equation}\label{eq:laplacehessian}
    p(\theta|D)\approx \mathcal{N}(\theta_{MAP}, H^{-1}).
\end{equation}

The computation of the Hessian term is still expensive. Therefore, further approximations are introduced in practice, most commonly the Generalized Gauss-Newton matrix \cite{martens2020new}. This takes the following form:
\begin{equation}
    H \approx \widetilde{H}=\sum_{n=1}^NJ_n^TH_nJ_n,
\end{equation}
where $J_n\in \mathcal{R}^{O\times W}$ is the Jacobian of the model outputs with respect to the parameters $\theta$ and $H_n\in\mathcal{R}^{O\times O}$ is the Hessian of the negative log-likelihood with respect to the model outputs, where $O$ denotes the model output size and $W$ the number of parameters. 

Given \eqref{eq:laplacehessian} during inference on unseen data, one cannot compute the full posterior predictive distribution but instead resort to sampling $\theta_s \sim p(\theta|D)$ for $s \in \{1, ...,S\}$ to approximate the predictions, 

\begin{equation}
    \hat{y}(x^{\star}) = \frac{1}{S} \sum_{s=1}^S f_{\theta_s}(x^{\star}),
\end{equation}

and obtain the predictive uncertainty by

\begin{equation}
    \sigma^2(x^{\star}) = \sqrt{\frac{1}{S} \sum_{s=1}^S f_{\theta_s}(x^{\star})^2 - \hat{y}(x^{\star})^2+\sigma^2}.
\end{equation}

For the subnet strategy, we include the options from the Laplace library for selecting the stochastic parameters.

\begin{table}[h!]
\centering
\resizebox{\columnwidth}{!}{%
\begin{tabular}{ l|c|l }
 \multicolumn{3}{c}{Summary of hyperparameters for the BNN with VI model} \\
 \hline
Hyperparameter & value range & hints  \\
 \hline
Number burn-in-epochs & $\approx[0, n]$ & after burn-in-epochs train variance and mean outputs. \\
Loss scale factor $\beta$ & $\beta \approx [100,500]$ & should depend on parameter and train set size. \\
Samples during training $S_{tr}$ & $S_{tr} \approx [5, 20]$ & depending on network size and computing resources. \\
Samples during tests and prediction $S_{te}$ & $S_{te} \approx [5, 50]$ & depending on network size and computing resources. \\
 \hline
\end{tabular}%
}
\label{table:hypers10}
\end{table}

\bigskip
\textbf{SWAG}:
Is an approximate Bayesian method and uses a low-rank Gaussian distribution as an approximation to the posterior over model parameters. The quality of approximation to the posterior over model parameters is based on using a high SGD learning rate that periodically stores weight parameters in the last few epochs of training \cite{maddox2019simple}. SWAG is based on Stochastic Weight Averaging (SWA), as proposed in \cite{izmailov2018averaging}. For SWA the weights are obtained by minimising the MSE loss with a variant of stochastic gradient descent. After, a number of burn-in epochs, $\tilde{t} = T-m$, the last $m$ weights are stored and averaged to obtain an approximation to the posterior, by

\begin{equation}\label{eq:swa}
    \theta_{SWA} = \frac{1}{m}\sum_{t=\tilde{t}}^T \theta_t.
\end{equation}

For SWAG we use the implementation as proposed by \cite{maddox2019simple}. Here the posterior is approximated by a Gaussian distribution with the SWA mean, \eqref{eq:swa} and a covariance matrix over the stochastic parameters that consists of a low rank matrix plus a diagonal, 

\begin{equation}\label{eq:swag}
    p(\theta |D) \approxeq \mathcal{N}\left(\theta_{SWA}, \frac{1}{2}(\Sigma_{diag}+\Sigma_{low-rank})\right).
\end{equation}

The diagonal part of the covariance is given by

\begin{equation}
    \Sigma_{diag} = \text{diag}(\bar{\theta^2} - \theta_{SWA}^2)
\end{equation}

where,

\begin{equation}\label{eq:swaggi}
    \bar{\theta^2} = \frac{1}{m}\sum_{t=\tilde{t}}^T \theta_t^2.
\end{equation}

The low rank part of the covariance is given by

\begin{equation}
    \Sigma_{low-rank} =  \frac{1}{m}\sum_{t=\tilde{t}}^T (\theta_t - \bar{\theta}_t) (\theta_t - \bar{\theta}_t)^T,
\end{equation}

where $\bar{\theta}_t$ is the running estimate of the mean of the parameters from the first $t$ epochs or also samples. In order to approximate the mean prediction, we again resort to sampling from the posterior \eqref{eq:swag}. With $\theta_s \sim p(\theta|D)$ for $s \in \{1, ...,S\}$, the mean prediction is given by

\begin{equation}
    \hat{y}(x^{\star}) = \frac{1}{S} \sum_{s=1}^S f_{\theta_s}(x^{\star}),
\end{equation}

and obtain the predictive uncertainty by

\begin{equation}
    \sigma(x^{\star}) = \sqrt{\frac{1}{S} \sum_{s=1}^S f_{\theta_s}(x^{\star})^2 - \hat{y}(x^{\star})^2}.
\end{equation}

For the subnet strategy, we include selecting the parameters to be stochastic by module names. 

\bigskip

\iffalse
\textbf{SGLD:} Stochastic gradient Langevin dynamics is an approximate sampling method, introduced in \cite{welling2011bayesian}. The posterior distribution over the weights is sampled by sampling from the parameter updates obtained by a variant of stochastic gradient descent. In SGLD Gaussian noise is injected into the parameter updates, such that the parameters $\theta$ do not collapse to just the MAP solution. The proposed update in \cite{welling2011bayesian} is

\begin{align}
    \Delta \theta_t &= \frac{\epsilon_t}{2} \left(\nabla \log p(\theta_t) + \frac{N}{n} \sum_{i=1}^n \nabla \log p(x_{ti}|\theta_t) \right) + \eta_t \nonumber \\
    \eta_t & \sim \mathcal{N}(0,\epsilon_t).
\end{align}

After, a number of burn-in epochs, $\tilde{t} = T-m$, the last $m$ weights are stored and averaged to obtain an approximation to the posterior. 
The mean prediction is then obtained as for a weighted ensemble,

\begin{equation}
    \hat{y}(x^{\star}) \simeq \frac{\sum_{t=\tilde{t}}^T \epsilon_t f_{\theta_t}(x^{\star})}{\sum_{t=1}^T \epsilon_t}.
\end{equation}

Another possibility is to resort to a simpler average as is usually done for MC sampling methods to obtain the mean prediction,

\begin{equation}
    \bar{y}(x^{\star})
     \simeq \frac{1}{m}\sum_{t=\tilde{t}}^T f_{\theta_t}(x^{\star}).
\end{equation}

The predictive uncertainty is then obtained as,

\begin{equation}
  \sigma(x^{\star}) = \sqrt{\frac{1}{m}\sum_{t=\tilde{t}}^T f_{\theta_t}(x^{\star})^2 - \bar{y}(x^{\star})^2}.
\end{equation}
\fi

\subsection{Gaussian Process Based UQ Methods}

\textbf{Recap of Gaussian Processes (GPs):} The goal of previously introduced methods was to find a distribution over the weights of a parameterized function i.e. a neural network. In contrast, the basic idea of a Gaussian Process (GP) is to instead consider a distribution over possible functions, that fit the data in some way. Formally,\newline

\textit{"A Gaussian process is a collection of random variables, any finite number of which have a joint Gaussian distribution."} \cite{seeger2004gaussian}\newline

Precisely, a GP can be described by a possibly infinite amount of function values
\begin{equation}
    f(x) \sim \mathcal{GP}(m(x),k_{\gamma}(x)),
\end{equation}

such that any finite collection of function values $f$ has a joint Gaussian
distribution,

\begin{align}
 f = f(X) = [f(x_1),\dots,f(x_K)]^{\top} \sim \mathcal{N}(m_X,\mathcal{K}_{X,X}) \,,  \label{eqn: gpdef}
\end{align}
with a mean vector, $(m_X)i = m(x_i)$, and covariance matrix, $(\mathcal{K}_{X,X})_{ij} = k_{\gamma}(x_i,x_j)$, 
stemming from the mean function $m$ and covariance kernel of the GP, $k_{\gamma}$, that is parametrized by $\gamma$. A commonly used covariance function is the squared exponential, also referred to as Radial Basis Function (RBF) kernel, exponentiated quadratic or Gaussian kernel:

\begin{equation}
    k_{\gamma}(x,x') = \text{cov}(f(x),f(x')) = \eta^2\exp{\left(-\frac{1}{2l^2}|x-x'|^2\right)}.
\end{equation}

Where $\gamma =( \eta^2, l)$ and $\eta^2$ can be set to $1$ or tuned as a hyperparameter. By default the lengthscale $l=1$ but can also be optimized over. Now the GP, $f (x) \sim GP(m(x), k(x, x'))$, as a distribution over functions can be used to solve a regression problem. Following \cite{seeger2004gaussian}, consider the simple case where the observations are noise free and you have training data $\mathcal{D}_{\mathrm{train}} = \{(x_i, y_i)\}_{i=1}^{N}$ with $X=(x_i)_{i=1}^N$ and $Y=(y_i)_{i=1}^N$. The joint prior distribution of the training outputs, $Y$, and the test outputs $f_*=f_*(X_*)= (f(i_k))_{i=1}^m$ where $X_* = (x_i)_{i=1}^m$ are the test points, according to the prior is

\begin{align}
 p(Y,f_*) = \mathcal{N}\bigg(0, \Bigg[ \begin{array}{rr}
\mathcal{K}_{X,X} & \mathcal{K}_{X,X_*}  \\
\mathcal{K}_{X_*,X} & \mathcal{K}_{X_*,X_*}  \\
\end{array} \Bigg] \bigg).
\end{align}

Here the mean function is assumed to be $m_X = 0$ and $\mathcal{K}_{X,X_*}$ denotes the $N \times m$ matrix of the covariances evaluated at all pairs of training and test
points, and similarly for the other entries $\mathcal{K}_{X,X}$, $\mathcal{K}_{X_*,X_*}$ and $\mathcal{K}_{X_*,X}$. To make predictions based on the knowledge of the training points, conditioning on the prior observations is used and yields,

\begin{align*}
    p(f_*|X_*, X, Y) & = \mathcal{N}(\mathcal{K}_{X_*,X}\mathcal{K}_{X,X}^{-1}Y, \mathcal{K}_{X_*,X_*}-\mathcal{K}_{X_*,X_*}\mathcal{K}_{X,X}^{-1}\mathcal{K}_{X,X_*})\\
    & = \mathcal{N}(m(X,X_*,Y), \tilde{\mathcal{K}}_{X,X_*}).
\end{align*}
  
Now to generate function values on test points, one uses samples from the posterior distribution $f_*(X_*) \sim \mathcal{N}(m(X,X_*,Y), \tilde{K}(X, X_*))$. To illustrate how we can obtain these samples from the posterior distribution, consider a Gaussian with arbitrary mean $m$ and covariance $K$, i.e. $f_* \sim \mathcal{N}(m,K)$. For this one can use a scalar Gaussian generator, which is available in many packages:

\begin{enumerate}
    \item  Compute the Cholesky decomposition of $K=LL^T$, where $L$ is a lower triangular matrix. This works because $K$ is symmetric by definition.
    \item Then, draw multiple $u \sim \mathcal{N}(0, I)$.
    \item Now, compute the samples with $f_* = m + Lu$. This has the desired mean, $m$ and covariance $L\mathbb{E}(uu^T)L^T = LL^T = K$.
\end{enumerate}

The above can be extended to incorporate noisy measurements $y \rightarrow y + e$, see \cite{seeger2004gaussian}, or noise on the inputs as in \cite{johnson2019accounting}. Both of these extensions require tuning of further hyperparameters, yet beneficially allow to incorporate a prediction of aleatoric uncertainty in a GP.\newline

For example, assume additive Gaussian noise on the distribution of the function values, 
\begin{equation}
    p(y(x)|f(x)) =  \mathcal{N}(y(x); f(x),\sigma^2).
\end{equation}
Then the predictive distribution of the GP evaluated at the
$K_*$ test points, $X_*$, is given by
\begin{align}
 p(f_*|X_*,X,Y,\gamma,\sigma^2) &= \mathcal{N}(\mathbb{E}[f_*],\text{cov}(f_*)) \,, \label{eqn: fullpred}  \\
 \mathbb{E}[f_*] &= m_{X_*}  + \mathcal{K}_{X_*,X}[\mathcal{K}_{X,X}+\sigma^2 I]^{-1}Y\,,   \notag \\
 \text{cov}(f_*) &= \mathcal{K}_{X_*,X_*} - \mathcal{K}_{X_*,X}[\mathcal{K}_{X,X}+\sigma^2 I]^{-1}\mathcal{K}_{X,X_*} \,.  \notag
\end{align}
Here $m_{X_*}$ is the $K_* \times 1$ mean vector, which is assumed to be zero in the previous case.\newline

In both cases, with and without additive noise on the function values, the GP is trained by learning interpretable kernel hyperparameters. The log marginal likelihood of the targets $y$ - the probability of the data conditioned only on kernel hyperparameters $\gamma$ - provides a principled probabilistic framework for kernel learning:
\begin{equation}
 \log p(y | \gamma, X) \propto -\left(y^{\top}(\mathcal{K}_{\gamma}+\sigma^2 I)^{-1}y + \log|\mathcal{K}_{\gamma} + \sigma^2 I|\right)\,,  \label{eqn: mlikeli}
\end{equation}
where $\mathcal{K}_{\gamma}$ is used for $\mathcal{K}_{X,X}$ given $\gamma$. Kernel learning can be achieved by optimizing Eq. \eqref{eqn: mlikeli} with respect to $\gamma$.\newline

The computational bottleneck for inference is solving the linear system
$(\mathcal{K}_{X,X}+\sigma^2 I)^{-1}y$, and for kernel learning it is computing
the log determinant $\log|\mathcal{K}_{X,X}+ \sigma^2 I|$ in the marginal likelihood.  
The standard approach is to compute the Cholesky decomposition of the
$K \times K$ matrix $\mathcal{K}_{X,X}$, which
requires $\mathcal{O}(K^3)$ operations and $\mathcal{O}(K^2)$ storage.
After inference is complete, the predictive mean costs $\mathcal{O}(K)$,
and the predictive variance costs $\mathcal{O}(K^2)$, per test point
$x_*$.

\bigskip
\textbf{Deep Kernel Learning (DKL)}: Conceptually DKL consists of a NN architecture that extracts a feature representation of the input $x$ and fits an approximate GP on top of these features to produce a probabilistic output \cite{wilson2016deep}. DKL combines GPs and DNNs in a scalable way. In practice, all parameters, the weights of the feature extractor and the GP parameters are optimized jointly by maximizing the log marginal likelihood of the GP. We utilize GPytorch for our implementation \cite{gardner2018gpytorch} and use a grid approximation where we optimized over the number of inducing points. For DKL the GP is transformed by replacing the inputs $x$ by the outputs of a NN in the following way. The kernel $k_{\gamma}(x,x')$ with hyperparameters
$\theta$ is replaced by,
\begin{equation}
k_{\gamma}(x,x') \to k_{\gamma}( g(x, \theta), g(x', \theta)) \,,  \label{eqn: deepkernel}
\end{equation}
where $g(x,\theta)$ is a non-linear mapping given by a deep architecture, such as a deep
convolutional network mapping into a feature space of dimension $J$, parametrized by weights $\theta$,

\begin{align}\label{eq:featurextract}
    g(\cdot, \theta): X &\rightarrow \mathbb{R}^J\nonumber \\
    x &\mapsto g(x, \theta).
\end{align}

This so called deep kernel in \eqref{eqn: deepkernel} is now used as the covariance function of a GP to model data $\mathcal{D} = \{x_i, y_i\}_{i=1}^{N}$. The deep kernel hyperparameters,
$\rho = \{ \gamma,\theta, \sigma^2 \}$, can be \emph{jointly} learned by maximizing the
log \emph{marginal likelihood} of the GP \eqref{eqn: mlikeliDNN}. 

\begin{equation}
 \mathcal{L} = \log p(Y | \gamma, X, \theta) \propto -\left(y^{\top}(K_{\gamma, \theta} +\sigma^2 I)^{-1}y + \log|K_{\gamma, \theta} + \sigma^2 I|\right)\,,  \label{eqn: mlikeliDNN}
\end{equation}

Except for the replacement of input data, one can almost follow the same procedures for learning
and inference as for GPs as outlined previously. For optimizing \eqref{eqn: mlikeliDNN} the chain rule is used to compute derivatives of the log marginal likelihood with respect to the deep kernel hyperparameters as in \cite{wilson2016deep}.\newline

 Exact inference is possible for the regression case, yet the computational complexity scales cubically with the number of data points and makes it not suitable for large datasets. Thus, following \citep{van2021feature} in the implementation the sparse GP of \citep{titsias2009variational} and the variational approximation of \citep{hensman2015scalable} is used, in order to allow for DKL to scale to large training datasets. The sparse GP approximation of \citep{titsias2009variational} augments the DKL model with $M$ inducing inputs, $Z \in \mathbb{R}^{M \times J}$, where $J$ is the dimensionality of the feature space, as in \eqref{eq:featurextract}. Moreover, to perform computationally efficient inference we use the the variational approximation introduced by \citep{hensman2015scalable}, where inducing points $Z$ are treated as variational parameters. $U$ are random variables with prior 
 
 \begin{equation}
     p(U) = \mathcal{N}(U | m_Z, \mathcal{K}_{Z,Z}),
 \end{equation}
and variational posterior
\begin{equation}
    q(U) =  \mathcal{N}(U | \widetilde{m}, S),
\end{equation}

where $\widetilde{m} \in \mathbb{R}^{M}$ and $S \in \mathbb{R}^{M \times M}$ are variational parameters and initialized at the zero vector and the identity matrix respectively.
 The approximate predictive posterior distribution at training points $X$ is then
\begin{equation}
  p(f| Y) \approxeq q(f) = \int p(f | U) q(U)  dU 
\end{equation}
Here $p(f | U)$ is a Gaussian distribution for which we can find an analytic expression, see \citep{hensman2015scalable} for details.
Note that we deviate from \citep{hensman2015scalable} in that our input points $X$ are mapped into feature space just before computing the base kernel, while inducing points are used as is (they are defined in feature space). The variational parameters $Z$, $\widetilde{m}$, and $S$ and the feature extractor parameters $\theta$ and GP model hyparparameters $\gamma$, given by $l$ and $\eta^2$, and $\sigma^2$ are all learned at once by maximizing a lower bound on the log marginal likelihood of the predictive distribution $p(Y|X)$, the ELBO, denoted by $\mathcal{L}$.
For the variational approximation above, this is defined as
\begin{equation}
  \log(p(Y|X)) \geq
  \mathcal{L}(Z, m,S, \gamma, \theta, \sigma^2) = \sum_{i=1}^N
  \mathbb{E}_{q(f)}
  \left[ \log p(y_i | f(x_i)) \right]
  - \beta \text{D}_{\text{KL}}(q(U) || p(U)).
\end{equation}

Both terms can be computed analytically when the likelihood is Gaussian and all parameters can be learned using stochastic gradient descent.
To accelerate optimization gpytorch additionally utilizes the whitening procedure of \citep{matthews2017scalable} in their Variational Strategy. The approximate predictive posterior distribution at test points $X^*$ is then
\begin{equation}
  p(f_* | Y) \approxeq q(f_*) = \int p(f_* | U ) q(U)  dU 
\end{equation}
For regression tasks we directly use the function values $f_*$ above as the predictions. 
We use the mean of $p(f_* | Y)$ as the prediction, and the variance as the uncertainty.

\section{Deterministic Uncertainty Estimation (DUE) - extension of DKL}

\begin{algorithm}[h!]
  \caption{Algorithm for training DUE \citep{van2021feature}}
  \label{alg:DUE}
  \begin{algorithmic}[1]
    \STATE \textbf{Definitions:}

    - Residual NN $g_{\theta}: x \rightarrow \mathbb{R}^J$ with feature space dimensionality J and parameters $\theta$.

    - Approximate GP with parameters $\rho = \{\gamma, \sigma^2, \omega\}$, where $\gamma = \{l, \eta\}$ and $l$ length scale and $\eta$ output scale of $k_{\gamma}$, $\omega$ GP variational parameters (including $m$ inducing point locations $Z$)

    - Learning rate $\zeta$, loss function $\mathcal{L}$
    \STATE Using a random subset of $p$ points of our training data, $X^\text{init} \subset X$, compute:

    \textbf{Initial inducing points:} K-means on $g_{\theta}(X^\text{init})$ with $K=m$. Use found centroids as initial inducing point locations Z in GP.

    \textbf{Initial length scale:} \\ $l = \frac{1}{\binom{p}{2}} \sum_{i=0}^p\sum_{j=i+1}^{p} |g_{\theta}(X_i^{\text{init}}) - g_{\theta}(X_j^{\text{init}})|_2$.

    \FOR{minibatch $x_b,y_b \subset X, Y$}
    \STATE $\theta' \leftarrow \text{spectral\_normalization}(\theta)$
    \STATE $p(y_b'|x_b) \leftarrow \text{evaluate\_GP}_\theta(g_{\theta'}(x_b))$
    \STATE $\mathcal{L} \leftarrow \text{ELBO}_\theta(p(y_b'|x_b), y_b)$
    \STATE $(\rho, \theta) \leftarrow (\rho, \theta) + \zeta * \nabla_{\rho, \theta} \mathcal{L}$
    \ENDFOR
  \end{algorithmic}
\end{algorithm}

 DUE builds on DKL by using the same model except for exchanging the feature extractor of the DKL model. With this replacement DUE addresses limitations of DKL and provides potentially robust uncertainty estimates. According to \citep{van2021feature} with DKL, data points dissimilar to the training data (also called OOD data) can potentially be mapped close to feature representations of in-distribution points. These feature representations, which are close in some norm, input into the approximate GP yield similar or nearly the same predictions. This is called "feature collapse", and suggests that a constraint must be placed on the deep feature extractor. Based on deterministic uncertainty quantification (DUQ) \citep{van2020uncertainty} and spectrally normaplized GPs (SNGP) \citep{liu2020simple}, the authors of \citep{van2021feature} propose to use a bi-Lipschitz constraint on a feature extractor.  This bi-Lipschitz constraint is enforced by spectral normalization on the weights, \citep{miyato2018spectral, gouk2021regularisation}.
This constraint mitigates so-called "feature collapse", by forcing the feature representation to be sensitive to changes in the input (lower Lipschitz, avoids feature collapse) but also generalize due to smoothness (upper Lipschitz).\newline

For convolutional and linear layers following \citep{van2021feature}, we use spectral normalization of the weight matrices to promote approximate bi-Lipschitz continuity. To promote spectral normalization for fully connected layers and $1 \times 1$ convolutions online power iteration are used and for larger convolutions an approximate method, as proposed in \citep{gouk2021regularisation} and was first implemented by \citep{behrmann2019invertible}, is used. Spectral normalization is also extended to batch normalization by rescaling the weights, see \citep{van2021feature} for details. 
Adding spectral normalization to batch normalization layers makes it more likely that the entire network's upper Lipschitz constant is bounded. The mean prediction and predictive uncertainty are obtained as for DKL.

\textit{Summary of learnable parameters:}
\begin{itemize}
    \item weights of DNN feature extractor $\theta$
    \item  for the GP, parameters $\gamma$: noise hyperparamter $\sigma^2$, the GP function mean $m$, the length scale of the GP kernel $l$ and the scale of the kernel $\eta^2$. In the above case the GP hyperparameters are learned by optimizing ELBO.
\end{itemize}

\textit{Summary of hyperparameters:}
\begin{itemize}
    \item number of power iterations for spectral normalization, usually set to $r=1$
    \item number of initial inducing points $M$
\end{itemize}

\subsection{Quantile Based UQ Methods}

\bigskip
\textbf{Quantile Regression (QR)}: The goal of Quantile Regression is to extend a standard regression model to also predict conditional quantiles that approximate the true quantiles of the data at hand. It does not make assumptions about the distribution of errors as is usually common. It is a more commonly used method in Econometrics and Time-series forecasting \cite{koenker1978regression}.\newline

In the following we will describe univariate quantile regression. Any chosen conditional quantile $\alpha \in [0,1]$ can be defined as 
\begin{equation}
    q_\alpha(x):=\text{inf}\{y \in \mathbb{R}: F(y \vert X = x ) \geq \alpha\},
\end{equation}

where $F(y \vert X = x)=P(Y\leq y\vert X = x)$ is a strictly monotonic increasing cumulative density function.\newline

 For Quantile Regression, the NN $f_{\theta}$ parameterized by $\theta$, is configured to output the number of quantiles that we want to predict. This means that, if we want to predict $p$ quantiles $[\alpha_1, ... \alpha_n]$, 
\begin{equation}
    f_{\theta}(x_{\star}) = (\hat{y}_1(x^{\star}), ...,\hat{y}_n(x^{\star})).
\end{equation}

The model is trained by minimizing the pinball loss function \cite{koenker1978regression}, given by the following loss objective,

\begin{equation}
    \mathcal{L}_i (\theta, (x^{\star}, y^{\star})) = \max\{ (1-\alpha_i) (y^{\star}-\hat{y}_i(x^{\star})), \alpha (y^{\star}-\hat{y}_i(x^{\star}))\}.
\end{equation}

Here $i \in \{1, ...,n\}$ denotes the number of the quantile and $100\alpha_i $ is the percentage of the quantile for $\alpha_i \in [0,1)$. Note that for $\alpha = 1/2$ one recovers the $\ell^1$ loss. During inference, the model will output an estimate for the chosen quantiles and these can be used as an indication of aleatoric uncertainty.

\bigskip
\textbf{Conformalized Quantile Regression (CQR)}: Conformal Prediction is a post-hoc uncertainty quantification method to yield calibrated predictive uncertainty bands with proven coverage guarantees \cite{angelopoulos2021gentle}. Based on a held out calibration set, CQR uses a score function to find a desired coverage quantile $\hat{q}$ and conformalizes the QR output by adjusting the quantile bands via $\hat{q}$ for an unseen test point as follows $x_{\star}$:

\begin{equation}
    T(x_{\star})=[\hat{y}_{\alpha/2}(x_{\star})-\hat{q}, \hat{y}_{1-\alpha/2}(x_{\star})+\hat{q}]
\end{equation}

where $\hat{y}_{\alpha/2}(x_{\star})$ is the lower quantile output and $\hat{y}_{1-\alpha/2}(x_{\star})$, \cite{romano2019conformalized}.\newline

\subsection{Diffusion Based UQ Methods}

\bigskip
\textbf{CARD}: The classification and regression diffusion (CARD) models, as introduced in \cite{han2022card}, combine a denoising diffusion-based conditional generative model and a pre-trained conditional mean estimator in order to obtain a predictive distribution given an input. Given a target $y^{\star}$ and input $x^{\star}$ CARD utilizes a series of intermediate predictions $y_{1:T}$ for a number of steps $T \in \mathbb{N}$. The parameters of the diffusion-based conditional generative model are obtained by optimising the following objective

\begin{equation}
    \mathcal{L}_{\text{ELBO}}(y^{\star}, x^{\star}) = \mathcal{L}_0(y^{\star}, x^{\star})+  \sum_{t=2}^T\mathcal{L}_{t-1}(y^{\star}, x^{\star})+  \mathcal{L}_T(y^{\star}, x^{\star}),
\end{equation}

where the individual terms are given by

\begin{equation}
    \mathcal{L}_0(y^{\star}, x^{\star}) = \mathbb{E}_q\left[-\log(p_{\theta}(y^{\star}|y_1,x)\right]
\end{equation}

\begin{equation}
    \mathcal{L}_{t-1}(y^{\star}, x^{\star}) = \mathbb{E}_q\left[D_{\text{KL}}(q(y_{t-1}|y_t,y_0,x) || p_{\theta}(y_{t-1}|y_t,x))\right]
\end{equation}

\begin{equation}
     \mathcal{L}_T(y^{\star}, x^{\star}) = \mathbb{E}_q\left[D_{\text{KL}}(q(y_T|,y_0,x) || p(y_T|x))\right]
\end{equation}

and the predictive distribution $p(y_T|x)$ is obtained by a MAP estimate, in our case the deterministic base model, 

\begin{equation}
    p(y_T|x) = \mathcal{N}(f_{\theta_{MAP}}(x),\mathbb{I}).
\end{equation}

Following \cite{pandey2022diffusevae} the forward process of conditional distributions with a diffision schedule $(\beta_t)_{t=1}^T \in (0,1)^T$ is defined such that a closed-form solution exists,

\begin{equation}
    p(y_t | y_{t-1}, f_{\theta_{MAP}}(x)) = \mathcal{N}(y_t; \sqrt{1-\beta_t} y_t + (1-\sqrt{1-\beta_t}) f_{\theta_{MAP}}(x), \beta_t \mathbb{I}),
\end{equation}

this admits a closed form and non-iterative solution at each time step $t \in \{1, ..., T\}$,

\begin{equation}
    p(y_t | y_0, f_{\theta_{MAP}}(x)) = \mathcal{N}(y_t; \sqrt{\alpha_t} y_t + (1-\sqrt{\alpha_t}) f_{\theta_{MAP}}(x), \beta_t \mathbb{I}),
\end{equation}

with $\alpha_t = \Pi_{l=1}^t (1-\beta_l)$. For regression the goal is to reverse the above diffusion process to recover the distribution of the noise term and, hence, obtaining the aleatoric uncertainty of the second moment predictive distribution $p(y|x)$. For this a neural network $\epsilon_{\theta}$ is trained that given a sample $y_t$ predicts the corresponding noise $\epsilon \approxeq \epsilon_{\theta}(x, y_t, f_{\theta_{MAP}}(x), t)$. The predictive mean and uncertainty, in terms of standard deviation, is obtained by moment matching with the predictive samples $y_0$ approximating the labels $y^{\star}$.

\subsection{Partially Stochastic Network Strategies}

In order to adapt the Bayesian UQ methods to large EO data sets, we support partially stochastic NNs following the approach presented in \cite{sharma2023bayesian}. In \cite{sharma2023bayesian} the authors demonstrate experimentally and theoretically that partially stochastic networks can also approximate predictive distributions. There are multiple ways to obtain partially stochastic networks. For the Laplace Approximation and SWAG methods, we use a two-stage training. First, all parameters are obtained by a MAP estimate. Then, in the second training stage the stochastic parameters are obtained.
For BNN with VI and BNN+LV we use joint training, where the stochastic and deterministic parameters are learnt jointly by maximising the evidence lower bound or the so called $\alpha$-divergence, \cite{depeweg2018decomposition}.

\section{Metrics}
\label{seq:metrics}

Regression tasks are commonly evaluated by accuracy metrics such as Root Mean Squared Error (RMSE) or coefficient of determination, $R^2$. A better quality of prediction is indicated by a lower RMSE and MAE and a R$^2$ score close to $1.0$.  However, these measures only characterize the error between point predictions and available targets. When considering UQ methods, we therefore need additional metrics in the form of proper scoring rules \cite{gneiting2007strictly} which do not ignore predictive uncertainty. In particular, we consider the negative log-likelihood (NLL) of a Gaussian as a proper scoring rule, \cite{gneiting2007strictly}. Moreover, we consider calibration as introduced in \cite{kuleshov2018accurate}. As neither the NLL or calibration are sufficient to verify a useful forecast since a model with large predictive uncertainties can be well calibrated and obtain a sufficient NLL, we additionally consider sharpness, which measures the mean of the predictive uncertainties. We use \cite{chung2021uncertainty} for metric computation and some plots. \\

The RMSE is computed between the targets $\boldsymbol{y} = (y_i)_{i=1}^N$ and the mean model predictions $\boldsymbol{f}(x) = (f(x_i))_{i=1}^N$ for $N$ samples as

\begin{equation}
    \text{RMSE}(\boldsymbol{f}(x),\boldsymbol{y}) = \sqrt{\frac{1}{N}\sum_{i=1}^N (f(x_i)-y_i)^2}.
\end{equation}

The MAE is computed as

\begin{equation}
    \text{MAE}(\boldsymbol{f}(x),\boldsymbol{y}) = \frac{1}{N}\sum_{i=1}^N |f(x_i)-y_i|.
\end{equation}

The R$^2$ is computed as

\begin{equation}
    \text{R}^2 = \text{R}^2(\boldsymbol{f}(x),\boldsymbol{y}) = 1 - \frac{\sum_{i=1}^N (f(x_i)-y_i)^2}{\sum_{i=1}^N \left(f(x_i)-\frac{1}{N}\sum_{j=1}^N f(x_j)\right)^2}.
\end{equation}

However, these measures only characterize the error between point predictions and available targets. In order to compare the predictive uncertainties to the target distribution, we need additional metrics, such as proper scoring rules \cite{gneiting2007strictly}. We consider the NLL of a Gaussian as a proper scoring rule \cite{gneiting2007strictly}. We also report the miscalibration area, where a lower miscalibration area indicates a better fit of the predictive uncertainties to the true target distribution. To quantify the overall confidence of a model in a single metric, we consider sharpness which computes the mean of the predictive uncertainties. We use \cite{chung2021uncertainty} for computing these metrics.

The NLL is computed between the targets $\boldsymbol{y} = (y_i)_{i=1}^N$ and the mean model predictions $\boldsymbol{f}(x) = (f(x_i))_{i=1}^N$ and predictive uncertainties $\boldsymbol{\sigma}(x) = (\sigma(x_i))_{i=1}^N$ for $N$ samples as NLL is computed as 

\begin{equation}
    \text{NLL}((\boldsymbol{f}(x), \boldsymbol{\sigma}(x)), \boldsymbol{y}) = \frac{1}{N}\sum_{i=1}^N\left(\frac{1}{2}\text{ln}\left(2\pi\sigma(x_i)^2\right) + \frac{1}{2\sigma(x_i)^2}\left(f(x_i)-y_i\right)^2\right),
\end{equation}

Additional we consider the scoring rule of the Continuous Ranked Probability Score (CRPS), which for single sample and a predictive distribution that is Gaussian is given by

\begin{equation}
    crps(\mathcal{N}(\mu, \sigma), y) = -\sigma \Bigg(\frac{y-\mu}{\sigma}(2\Phi\bigg(\frac{y-\mu}{\sigma}\bigg)-1)+2\phi\bigg(\frac{y-\mu}{\sigma}\bigg)-\frac{1}{\sqrt{\pi}}\Bigg),
    \end{equation}
    
where $\Phi$ is the cumulative density function and $\phi$ probability distribution of $\mathcal{N}(0,1)$. Then, we compute the average sum over all predictions and labels, where $f_{\theta}(x^\star_i) = (\mu(x^\star_i), \sigma(x^\star_i))$,  which gives the reported CRPS,

\begin{equation}
    CRPS = \frac{1}{N^\star} \sum_{i=1}^{N^\star} crps(f_{\theta}(x^\star_i), y^\star_i).
\end{equation}

The miscalibration area is computed based on \cite{chung2021uncertainty} and is identical to mean absolute calibration error, however the integration here is taken by tracing the area between curves.

The sharpness is computed as

\begin{equation}
    \text{sharpness}(\boldsymbol{\sigma}(x)) = \sqrt{\frac{1}{N}\sum_{i=1}^N\sigma(x_i)^2}.
\end{equation}

Another key aspect for assessing the reliability of uncertainty estimates is calibration. Calibration refers to the degree to which a predicted distribution matches the true underlying distribution of the data. The mean absolute calibration error,  (MACE), gives the mean absolute error of the expected and observed proportions for a given range of quantiles.

\section{Training Details}

 We train all methods with the SGD or Adam optimizer \cite{kingma2014adam} with default parameters for a minimum of $50$ epochs until convergence based on the validation loss and evaluate the trained model on the in and out of distribution sets. BNNs, DKL, DUE and CARD were trained with Adam while other methods where trained with SGD.
 All methods and experiments are implemented in Pytorch \cite{paszke2019pytorch} and Lighting \cite{falcon2019pytorch} to enhance reproducibility of results. \\ For the Deep Ensemble we use five independently trained Gaussian Networks, for MC-Dropout and SWAG we use the settings suggested by \cite{maddox2019simple}, for the Laplace Approximation we use a Kronecker factored Hessian approximation through the Laplace library \cite{daxberger2021laplace}, for the DKL implementation we follow \cite{van2021feature}, for the BNN we use \cite{krishnan2022bayesiantorch} with their default parameters and for Quantile Regression we predict quantiles $0.1$, $0.5$, and $0.9$ \cite{angelopoulos2021gentle}.

\bibliography{iclr2024_conference}
\bibliographystyle{iclr2024_conference}

%% file: math_commands.tex
\usepackage{amsmath,amsfonts,bm}

\def\eqref#1{equation~\ref{#1}}
\def\1{\bm{1}}

\DeclareMathAlphabet{\mathsfit}{\encodingdefault}{\sfdefault}{m}{sl}
\SetMathAlphabet{\mathsfit}{bold}{\encodingdefault}{\sfdefault}{bx}{n}